\documentclass[twocolumn,prb]{revtex4-1}
\usepackage[utf8]{inputenc}
\usepackage{amsmath,amsfonts,amssymb,graphicx,multirow,overpic,dsfont,bm,braket,mathtools,xcolor,tikz,soul}
\usepackage{blkarray}

\begin{document}
\title{
Strong planar subsystem symmetry-protected topological phases \protect\\ and their dual fracton orders
}
\author{Trithep Devakul}
\affiliation{Department of Physics, Princeton University, Princeton, NJ 08540, USA}
\author{Wilbur Shirley}
\affiliation{Department of Physics and Institute for Quantum Information and Matter, California Institute of Technology, Pasadena, California 91125, USA}
\author{Juven Wang}
\affiliation{Center of Mathematical Sciences and Applications, Harvard University,  Cambridge, MA 02138, USA}
\affiliation{School of Natural Sciences, Institute for Advanced Study,  Einstein Drive, Princeton, NJ 08540, USA}
\date{\today}

\begin{abstract}
We classify subsystem symmetry-protected topological (SSPT) phases in 3+1D protected by planar subsystem symmetries: 
short-range entangled phases which are dual  to long-range entangled abelian fracton topological orders via a generalized `gauging' duality.
We distinguish between weak SSPTs, which can be constructed by stacking 2+1D SPTs, and strong SSPTs, which cannot.
We identify signatures of strong phases, and show by explicit construction that such phases exist.
A classification of strong phases is presented for an arbitrary finite abelian group.
Finally, we show that fracton orders realizable via $p$-string condensation are dual to weak SSPTs,
while those dual to strong SSPTs exhibit novel statistical interactions prohibiting such a realization.
\end{abstract}

\newcommand{\ve}{\mathbf}
\newcommand{\matr}{\mathbf}

\maketitle

\emph{Introduction---}
Global symmetries, such as the $\mathbb{Z}_2$ spin-flip symmetry of the Ising model, act throughout the bulk of a system.
Recently, there has been an emerging interest in symmetries that act on only \textit{part} of a system.
These include higher-form symmetries which act on deformable lower-dimensional manifolds of a system~\cite{higherform},
as well as subsystem symmetries~\cite{subsys1,subsys2,subsys3}, which act on rigid lower-dimensional subsystems.
It has also been realized that such subsystem symmetries may protect non-trivial symmetry-protected topological (SPT) phases~\cite{spt1,spt2,groupcohomology,spt3}: gapped, disordered, short-range entangled phases
which cannot be adiabatically connected to the trivial disordered phase in the presence of symmetry, but can be if the symmetry is not enforced.
Examples of subsystem symmetries include those which act along linear~\cite{you1dsspt,dwy},  planar~\cite{vijayhaahfugauge,you2dsspt}, or even fractal~\cite{yoshidafrac1,devakulfracsspt,devakulfracclass,williamsongauge,newmanmoore,castelnovo} subsystems.
Such phases have been aptly named subsystem SPT (SSPT) phases, and this paper concerns their classification.

In 2+1D, such systems have gained interest due to the discovery that non-trivial SSPT phases may serve as a resource for universal measurement-based quantum computation (MBQC)~\cite{clusterstate,mbqcclus,mbqcsspt1,mbqcsspt2,mbqcfrac,mbqcssptarch} 
and also due to their unusual patterns of quantum entanglement~\cite{williamsonspee,dwy,schmitz,stephenspee,haahspee}.
In attempting to classify 2+1D linear SSPTs, one is faced with the issue that 
there are uncountably infinitely many distinct phases.
This is due to the presence of \textit{weak} phases: SSPT phases which can be constructed by stacking (a process which we will define) 1+1D SPTs along the subsystems, whose nontriviality are simply a manifestation of lower dimensional physics.
Ref.~\onlinecite{dwy} defined an equivalence relation between phases wherein two phases that differ by stacking 1+1D SPTs belong to the same equivalence class.
Phases not in the trivial equivalence class are, by definition, \textit{strong} SSPTs.
It was found that there are a small number of equivalence classes, which provided a sensible classification for the uncountably infinite phases.
In contrast, note that for 2+1D fractal SSPTs, weak phases do not exist and the number of phases is countably infinite~\cite{devakulfracclass}.
This paper is the natural extension of Ref.~\onlinecite{dwy} to planar symmetries in 3+1D (henceforth, simply 3D). 

Systems with planar subsystem symmetries have also received intense interest recently due to the discovery that, under a generalized `gauging' duality~\cite{vijayhaahfugauge,williamsongauge,shirleygauge}, they map on to long-range entangled models exhibiting \textit{fracton} topological order~\cite{chamonfracton,bravyifracton,haahfracton,yoshidafrac1,vijayfracton,nandkishore,pai,pretkoelastic,pretkosupersolid,radzihovsky,prem,prem2,pretko1,pretko2,prem3,bulmash,slaglekim,coupledlayer,youlitinski,song}.
An example of such a system is the plaquette Ising model~\cite{johnston,vijayhaahfugauge,castelnovo2}, whose paramagnetic phase is dual to the X-cube model of fracton topological order~\cite{vijayhaahfugauge}.
Fracton phases are characterized by a subextensive topological ground state degeneracy growing exponentially with $L$, and quasiparticle excitations with limited mobility.
The classification of such fracton phases is an active topic of research~\cite{gromovclass,duasort,paihermele,foliated,foliated2,foliated3,foliated4,foliated5,bifurcating}.
In this paper, we focus on classification of SSPT phases which are dual to abelian fracton phases, thus also providing a useful means of categorizing such fracton phases.

The brief history of 3D planar SSPT phases begins with Ref.~\onlinecite{you2dsspt}, which constructed a non-trivial 3D planar SSPT model.
However, it was later discovered that its fracton dual belonged to the same foliated fracton phase as the X-cube model~\cite{foliated6}, implying that it is weak.
More recently, fracton phases were constructed in Ref.~\onlinecite{foliated5} which possess `twisted' foliated fracton orders, raising the question as to the nature of their SSPT duals.
We find that these phases, too, are weak.
This prompts the question: do any strong planar SSPTs exist?
We answer this in the affirmative.
We explicitly construct strong SSPT phases, which are dual to novel strong fracton phases with unusual braiding statistics that cannot be obtained by coupling 2D theories.
In this sense these statistical interactions are ``intrinsically" three-dimensional.

We will first show how to construct weak 3D planar SSPT phases via a stacking process of 2D SPTs.
We then ask whether there are SSPT phases which cannot be realized by this process.
We identify mechanisms by which an SSPT may be strong, leading to a classification of such phases, and construct exactly solvable, zero-correlation length models realizing these phases.
In the fracton dual picture, this construction corresponds to one in which 2D topological orders are stacked on to and strongly coupled to an existing fracton model~\cite{foliated5}.
The duals of our strong SSPTs are novel fracton phases which cannot be attained via such a procedure, also implying that they cannot be realized by a $p$-string condensation transition~\cite{coupledlayer,coupledlayer2}, as we will show.

\emph{Planar subsystem symmetries---}
Throughout we will consider a system with degrees of freedom on each site of a cubic lattice.
Each site $\ve{r}$ transforms under the finite abelian on-site  symmetry group $G$ under a unitary representation $u_{\ve{r}}(g)$, where $g\in G$.
An $xy$ planar symmetry acting on plane $z$ acts as $S^{xy}(z;g) = \prod_{x,y} u_{\ve{r}=(x,y,z)}(g)$ for $g\in G$.  
Similarly, we may define $S^{yz}(x;g)$ and $S^{zx}(y;g)$, which act on $yz$ and $zx$ planes respectively.
Importantly, individual sites transform under the same on-site representation regardless of the orientation of the planar symmetry --- 
there is therefore a redundancy: the product of all $xy$ symmetries is identical to the product of all $yz$ or all $zx$ symmetries.
We will refer to models which respect only one orientation of planar symmetry as $1$-foliated, those with two as $2$-foliated, and those with all three as $3$-foliated.
To construct explicit models, we choose the on-site degrees of freedom to be $G$-valued, $\ket{g_\ve{r}}$, which transform under the on-site symmetry as $u_\ve{r}(g)\ket{g_\ve{r}} = \ket{g g_\ve{r}}$.

\emph{Construction of weak SSPT phases---}
It is possible to construct non-trivial SSPT phases from known 2D global SPTs, as we will show in this section.
Phases obtained in this way are `weak', by definition, whose nontrivial properties are in some sense a manifestation of lower-dimensional physics.
We emphasize here that we do not assume any translation invariance in our system.
Hence, our definition is different (but similar in spirit) to weak crystalline SPTs with global symmetries, which are stacks of lower dimensional SPTs protected by translation symmetry.

First, we briefly review the group cohomological classification of 2D bosonic SPTs with global symmetry group $G$~\cite{groupcohomology,levingu}.
For the purpose of being self-contained, we also include a more detailed review in the Supplementary Material~\cite{supp}.
The classification of such phases~\cite{groupcohomology,elsenayak} is given by the third cohomology group $H^3[G,U(1)]$.
For simplicity, we may consider $G=(\mathbb{Z}_N)^n$,
in which case an element of $H^3[G,U(1)]$ is specified by integers, $p_I^{i}$ ($i\in[1,n]$), $p_{II}^{ij}$ ($i<j$), and $p_{III}^{ijk}$ $(i<j<k)$, all modulo $N$, called type-I, II, and III cocycles respectively. 
We will specify $p_{I}^i$ and $p_{II}^{ij}$ compactly in a single symmetric $n\times n$ integer matrix $\matr{M}$ with $M_{ii} = 2 p_I^i$ and $M_{ij}=M_{ji}=p_{II}^{ij}$.
Upon gauging the global symmetries of a 2D SPT, one obtains a topologically ordered system with fractional quasiparticles carrying gauge charge or flux (or both). 
Nontrivial type-III cocycles give rise to non-abelian topological order,~\cite{propitius,wanggc1} which we will not consider here.
A generating set of quasiparticles are the ``electric'' excitations (gauge charges) $\{e_i\}$ and ``magnetic'' excitations (gauge fluxes) $\{m_i\}$.
Each $e_i$ has a $e^{2\pi \mathrm{i} / N}$ braiding statistic with $m_i$ and trivial statistics with all other generators.
The elements of $\matr{M}$ characterize the self and mutual statistics of gauge flux excitations~\cite{levingu}.
In particular, the type-I cocycles give rise to a self exchange statistic $e^{{\pi \mathrm{i} M_{ii}}/{N^2}}$ of the gauge flux $m_i$, and type-II cocycles lead to a mutual braiding statistic of $e^{{2\pi \mathrm{i} M_{ij}}/{N^2}}$ between $m_i$ and $m_j$. 
Note that these phases are only well defined modulo $e^{2\pi \mathrm{i}/N}$, since flux is only well defined up to attachment of charge, $m_i\rightarrow m_i e_j$.
Finally, we note that abelian topological orders in $2$D can all be described by $\matr{K}$ matrix Chern-Simons theories.~\cite{wenzee,wen2dtheory}
The topological orders we have discussed have a $2n\times 2n$ $\matr{K}$ matrix description with
\begin{equation}
\matr{K} = N
\begin{bmatrix}
-\frac{1}{N}\matr{M} & \matr{1}\\
\matr{1} & 0
\end{bmatrix}
,\;\;\;\;
\matr{K}^{-1} = \frac{1}{N}
\begin{bmatrix}
0 & \matr{1}\\
\matr{1} & \frac{1}{N} \matr{M}
\end{bmatrix}
\end{equation}
where the indices labeling quasiparticles are ordered as $\{e_1,\dots, e_n, m_1, \dots, m_n\}$.
Quasiparticles are described by an integer vector ${\bf \ell}$ in this basis, and
have self-exchange statistic {$e^{\pi \mathrm{i} {\bf \ell}^T\cdot \matr{K}^{-1}\cdot {\bf \ell}}$}
and mutual braiding statistics {$e^{2\pi \mathrm{i} {\bf \ell}_1^T\cdot \matr{K}^{-1}\cdot {\bf \ell}_2}$}.

\begin{figure}
    \includegraphics[width=0.5\textwidth]{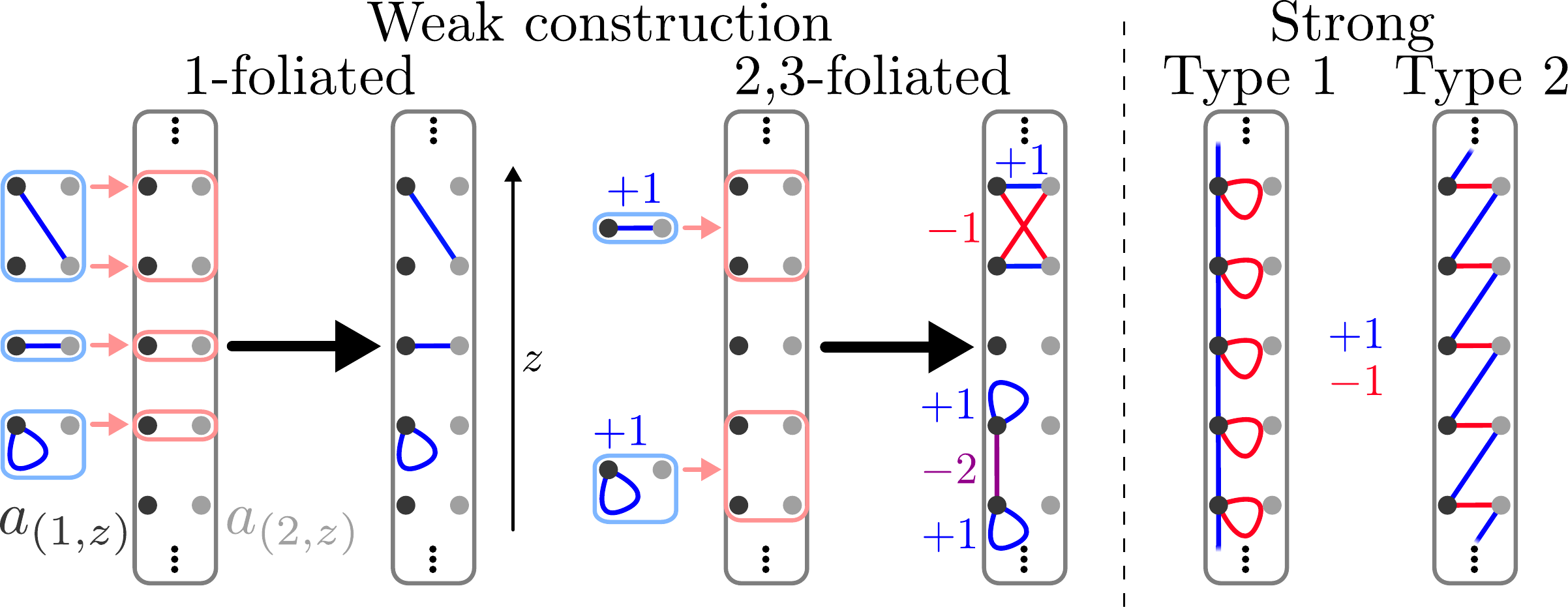}
    \caption{
    (Left) Examples of our construction of 1-foliated or weak 2 or 3-foliated models, for $G=\mathbb{Z}_N\times\mathbb{Z}_N$, in the graphical notation.
    2D SPTs to be stacked,
    are shown in the blue boxes, and the large arrow points to the resulting SSPT after stacking.
    The color of the edges connecting two vertices indicate its weight modulo $N$.
    (Right) Examples of $\matr{M}$ matrices that cannot be obtained by stacking 2D phases onto 2 or 3-foliated models.
    The Type 1 phase is only strong for even $N$, and Type 2 strong phases can only be realized for 2-foliated symmetries.
    }\label{fig:graph}
\end{figure}
It is always possible to view a 3D planar SSPT as a quasi-2D system in the $xy$ plane with a subextensively large symmetry group $G^{L}$ by compactifying the $z$ direction.
We may then proceed to compute its classification in terms of $H^3[G^L,U(1)]$, which is characterized by a subextensively large $\matr{M}$ matrix. 
We note that it is possible to define $\matr{M}$ matrices corresponding to $yz$ or $zx$ as well, but for reasons that will become clear we will always consider the $xy$ symmetries only.
It is useful to introduce a graphical notation for $\matr{M}$, which is used in Fig.~\ref{fig:graph}.
The $\alpha$th generator of $G$ in a plane $z$ is denoted by a vertex $a_{i=(\alpha,z)}$.
Two vertices $i$ and $j$ are connected by an undirected edge with weight $M_{ij}$, and a vertex $i$ is connected to itself via a self-loop with weight $M_{ii}/2$, where weights are defined modulo $N$.

Consider the 2D global symmetry group $G_{2D}=G^K$ for an integer $K$. For appropriate choice of the pure phase function $f_{2D}$, the wavefunction
$\ket{\psi}_{2D} = \sum_{\{g_\ve{r}\}} f_{2D}(\{g_\ve{r}\})\ket{\{g_\ve{r}\}}_{2D}$
on a 2D square lattice
is a zero-correlation length ground state of a commuting Hamiltonian with SPT order~\cite{supp}. 
All phases in the group cohomology classification can be realized in this way~\cite{Wang1403.5256,wanggc1,wanggc2}.

Suppose we start with the trivial disordered wavefunction  $\ket{\psi_0}=\sum_{\{g_\ve{r}\}} \ket{\{g_\ve{r}\}}$ on the 3D cubic lattice.
We can construct a nontrivial 1-foliated SSPT by identifying each factor of $G$ in $G_{2D}$ in the function $f_{2D}(\{g_r\})$ with a planar $G$ symmetry in an arbitrary collection of planes $z_1,\dots,z_K$ (where $z_k$ are all within some finite range to ensure locality).
The wavefunction $\ket{\psi}_\textrm{1-fol} = U\ket{\psi_0}$ 
with $U=\sum_{\{g_\ve{r}\}} f_{2D}(\{g_\ve{r}\}_{r_z\in\{z_k\}}) \ket{\{g_\ve{r}\}}\bra{\{g_\ve{r}\}}$
is the ground state of a 1-foliated 3D SSPT, which is nontrivial only near the planes $z_k$.
We may then repeat this procedure arbitrarily many times, each time acting on the previous state with $U$ for different choices of $f_{2D}$ and $\{z_k\}$.
We will call this procedure ``stacking'' the 2D SPT $\ket{\psi}_{2D}$ onto the planes $\{z_k\}$ of a 1-foliated SSPT.

More generally, we may define a stacking operation between two SSPTs in which the two systems, with on-site symmetry representations $u_\ve{r}^{(1)}(g)$ and $u_\ve{r}^{(2)}(g)$, are placed on top of each other to create a new SSPT with on-site representation $u_\ve{r}(g) = u_\ve{r}^{(1)}(g)\otimes u_\ve{r}^{(2)}(g)$.
The group structure of the standard SPT classification is realized under such a stacking operation.
Stacking a 2D SPT onto a 3D SSPT can be viewed as stacking two 3D SSPTs, in which the first is only nontrivial in the vicinity of a number of planes $\{z_k\}$. 
We define any phase realizable by stacking 2D SPTs in this way to be weak.
In the case of our 1-foliated SSPT construction, each additional stacked 2D SPT simply adds to the corresponding elements of $\matr{M}$, shown graphically in Fig.~\ref{fig:graph}.
For 1-foliated symmetries, it is thus possible to realize any $\matr{M}$ by stacking 2D SPTs; hence all phases are weak.

On the other hand, for $2$- or $3$-foliated models, this procedure may not work because $\ket{\psi}_\textrm{1-fol}$ is not guaranteed to be symmetric under the orthogonal planar symmetries (if it is, we can simply follow the same procedure).
Instead, let us define variables $d_{\ve{r}} = g_{\ve{r}+\ve{z}}g_\ve{r}^{-1}$, which transform under $xy$ planar symmetries 
but are invariant under all orthogonal symmetries.
We may then define non-trivial SSPT wavefunctions as before, but in terms of $d_\ve{r}$ instead using  the unitary
\begin{equation}
U = \sum_{\{g_\ve{r}\}} f_{2D}(\{d_\ve{r}\}_{r_z\in\{z_k\}})\ket{\{g_\ve{r}\}}\bra{\{g_\ve{r}\}},\label{eq:Udr}
\end{equation}
which is explicitly invariant under the orthogonal symmetries.  
However, in this case the $\matr{M}$ matrix of the 2D SPT does not map directly onto that of the SSPT --- instead one should view the 2D SPT as living ``in between'' the planes of the SSPT, at $\{z_k+1/2\}$.
To obtain the $\matr{M}$ matrix of the SSPT, one can compute the appropriate type-I and II cocycles of the 2D SPT in the basis of the $xy$ planar symmetries~\cite{supp}.
This process is shown in Fig.~\ref{fig:graph}.
As will be discussed in the next section, unlike for $1$-foliated symmetries,
there are now allowable phases which cannot be realized by stacking any number of 2D SPT.

Note that in this discussion we have implicitly ignored nontrivial SSPTs that have trivial $\matr{M}$ matrices. 
Such phases do exist~\cite{supp}.
However, we conjecture that all such phases are weak (they can be realized by stacking 2D linear SSPTs~\cite{dwy}) and therefore irrelevant in the classification of strong phases.

\emph{General constraints and invariants---}
In the presence of orthogonal symmetries, there are general constraints that must be satisfied by $\matr{M}$.
Conceptually, these arise due to the aforementioned redundancy: the global symmetry $S_{\mathrm{glob}}(g)=\prod_z S^{(xy)}(z;g)=\prod_x S^{(yz)}(x;g)$.
Since $yz$ symmetries do not contribute to $\matr{M}$, the generator $S_\mathrm{glob}(g)$ must therefore manifest trivially in $\matr{M}$.
This leads to two types of constraints on the elements of $\matr{M}$: 
the global symmetry must have trivial  type-II cocycle with any other symmetry and trivial type-I cocycle with itself.
We prove that these constraints must hold generally~\cite{supp}.
Let us label the $\alpha$th generator of $G$ on the $z$th plane by $i=(\alpha,z)$.
Then, the two constraints are expressed as
\begin{align}
\begin{split}
\sum_{z^\prime}^{} M_{(\alpha,z),(\beta,z^\prime)} \equiv 0
\mod N, \;\;\forall \alpha,z,\beta
\end{split}\label{eq:localconstraint}
\end{align}
and 
\begin{align}
\frac{1}{2}\sum_{z,z^\prime}^{}  M_{(\alpha,z),(\alpha,z^\prime)} \equiv 0
\mod N,
\;\;\forall \alpha
\end{align}
 
These constraints define a restricted subgroup of $H^3[G^L,U(1)]$ in which $2$- or $3$-foliated SSPTs must reside.
As we will show, there are now allowed phases which cannot be realized by stacking any number of 2D SPTs --- these are precisely the strong phases we are searching for.
This motivates us to define two types of strong invariants, $F_1$ and $F_2$, which cannot be changed by stacking with 2D SPTs.

\emph{Strong SSPTs: Type 1---}
Consider $G=\mathbb{Z}_{2N}$. Then $M_{z z^\prime}$ is an $L\times L$ matrix.  
Pick an arbitrary cut that divides the system into two halves $z<z_0$ and $z\geq z_0$.
Then,
\begin{equation}
F_1\equiv \sum_{z<z_0} \sum_{z^\prime \geq z_0} M_{z z^\prime} \mod 2
\end{equation}
is a $\mathbb{Z}_2$-valued global invariant.
To see why, view $M_{z z^\prime}\;\mathrm{mod}\;2$ as a $\mathbb{Z}_2$ ``flux'' flowing from vertex $z$ to $z^\prime$ in the graphical representation.  
Then, Eq.~\ref{eq:localconstraint} is a divergence-free constraint at each vertex.
The invariant $F_1$ is simply the total $\mathbb{Z}_2$ flux flowing through a cut at $z_0$.
It is therefore clear that $F_1$ does not depend on the choice of cut $z_0$, nor can it be modified by stacking a 2D SPT which amounts to adding closed flux loops locally.

\emph{Type 2---}
Consider $G=\mathbb{Z}_N\times\mathbb{Z}_N$, so that $M_{(\alpha,z),(\beta,z^\prime)}$ is a $2L\times 2L$ matrix.
Again pick a cut $z_0$.  Then,
\begin{equation}
F_2 \equiv\sum_{z<z_0}\sum_{z^\prime\geq z_0} \left(M_{(1,z),(2,z^\prime)} - M_{(2,z),(1,z^\prime)}\right)\mod N
\end{equation}
is a $\mathbb{Z}_N$-valued global invariant.
To see how this arises, interpret $M_{(1,z),(2,z^\prime)}$ as a $\mathbb{Z}_N$ ``flux" flowing from vertex $(1,z)$ to $(2,z^\prime)$.  
Like before, Eq.~\ref{eq:localconstraint} is a divergence-free constraint on this flux and $F_2$ measures the total flux flowing across a cut, which therefore does not depend on $z_0$ nor can it be modified by stacking with 2D SPTs.

In the Supplementary Material~\cite{supp}, we prove three important statements.
First, that the invariant $F_1$ or $F_2$ is the same regardless of whether we consider the $\matr{M}$ matrix obtained from $xy$ symmetries or that obtained from $yz$ (or $zx$) symmetries.
Secondly, 3-foliated systems must have trivial $F_2=0$.  
Thirdly, the set of $F_1$ and $F_2$ (which we also define for general $G$) completely classify $\matr{M}$ modulo stacking with 2D SPTs.
Finally, we also provide an explicit construction of a 3-foliated model which realizes a non-trivial type 1 strong phase $F_1=1$, and a 2-foliated model which realizes arbitrary $F_1$ and $F_2$, thereby demonstrating the existence of such strong phases. 
Examples of $\matr{M}$ matrices with non-trivial $F_1$ and $F_2$ are shown in Fig.~\ref{fig:graph} (right).

Let us define a `strong' equivalence relation between SSPTs, under which two phases belong to the same equivalence class if they can be connected with one another by stacking of 2D phases (along with, of course, symmetric local unitary transformations and addition/removal of disentangled degrees of freedom transforming as an on-site linear representation of $G$~\cite{Xie}).
For an arbitrary finite abelian group $G$, the set of equivalence classes is given by
\begin{align}
    C_\textrm{3-fol}[G] &= \prod_i \mathbb{Z}_{\gcd(2,N_i)}\\
    C_\textrm{2-fol}[G] &= \prod_i \mathbb{Z}_{\gcd(2,N_i)} \times \prod_{i<j} \mathbb{Z}_{\gcd(N_i,N_j)}
\end{align}
for 3-foliated and 2-foliated models respectively.
The group structure is realized via the stacking operation between two SSPTs.
We note that this equivalence relation can be naturally formulated in terms of planar-symmetric local unitary circuits, generalizing the definition of Ref.~\onlinecite{dwy}. Indeed the unitaries $U$ used to construct weak SSPTs are examples of such circuits.

\emph{Fracton duals---}
It is well known that, under a generalized gauge duality~\cite{vijayhaahfugauge,williamsongauge,shirleygauge}, SSPT phases map onto models of fracton topological order~\cite{you2dsspt,foliated5}.
The simplest and most well-studied fracton model is the X-cube model~\cite{vijayhaahfugauge}, which is obtained by gauging the planar symmetries of the plaquette Ising paramagnet,
and hosts fractional quasiparticle excitations with limited mobility including immobile fractons, lineons mobile along lines, and planons mobile within planes (which are either fracton dipoles or lineon dipoles).
For our discussion, we will assume that the reader has a rudimentary understanding of the X-cube fracton model and its quasiparticle excitations (see Ref.~\onlinecite{nandkishore} for a review).

Let us begin with 3-foliated SSPTs, which are dual to `twisted' X-cube fracton topological orders with fractonic charge~\cite{foliated5}.
The gauge flux $m_{(g,z)}$ of an element $g$ on the plane $z$ is a planon:
a composite excitation composed of a lineon anti-lineon pair on the planes $z+1/2$ and $z-1/2$, i.e. a lineon dipole.
A single lineon can be regarded as a semi-infinite stack of lineon dipoles mobile in the $x$ and $y$ directions.
For a more nuanced discussion of the mobility of such excitations, see Supplementary Material.

The constraints on the matrix $\matr{M}$ have a simple interpretation in this language:
the infinite stack of lineon dipoles, which belongs to the vacuum superselection sector~\cite{honeycombmodel}, must have trivial braiding statistics with all other lineon dipoles, and a trivial exchange statistic with itself.
The invariant $F_1$ also has a simple interpretation in this picture:
the quantity $e^{2\pi \mathrm{i} F_1/N^2}$ corresponds to the braiding (or crossing~\cite{coupledlayer}) statistic of a lineon and its anti-lineon on the same plane, modulo $e^{{4\pi \mathrm{i}}/{N^2}}$.

It is possible to construct fracton topological orders by strongly coupling intersecting stacks of topologically ordered 2D discrete gauge theories oriented along the $xy$, $yz$, and $zx$ planes, inducing a type of transition called $p$-string condensation~\cite{coupledlayer,coupledlayer2}.
More generally, these stacks of 2D gauge theories can be replaced by arbitrary 1-foliated gauge theories~\cite{foliated5}.
The twisted X-cube models that emerge from this construction are dual to weak 3-foliated SSPTs constructed via the planar-symmetric local unitaries $U$ in Eq.~\ref{eq:Udr}.
We walk through this correspondence in more detail in the Supplementary Material~\cite{supp}.

Equivalently, twisted X-Cube models dual to weak SSPTs may be obtained by effectively ``binding'' 2D anyons to existing planons in the fracton model.
As an example, consider placing one layer of the doubled semion topological order (with bosonic $e$ and semionic $m$)
onto a plane $z_0$ of the X-Cube model, and condensing pairs of $e$ and fracton dipoles in the plane $z_0$.
The end result is that $x$ or $y$ mobile lineons on plane $z_0$ and $m$ become confined, but the bound state of the two remain deconfined and form the new lineon excitations.
Since $m$ is a semion, the new lineons now also inherit their semionic statistics.
This procedure can be extended to general twisted quantum doubles living on multiple planes $\{z_k\}$, thereby binding more general 2D anyons to the lineons; this process is exactly dual to stacking a 2D SPT according to Eq.~\ref{eq:Udr}.

\begin{figure}
    \includegraphics[width=0.5\textwidth]{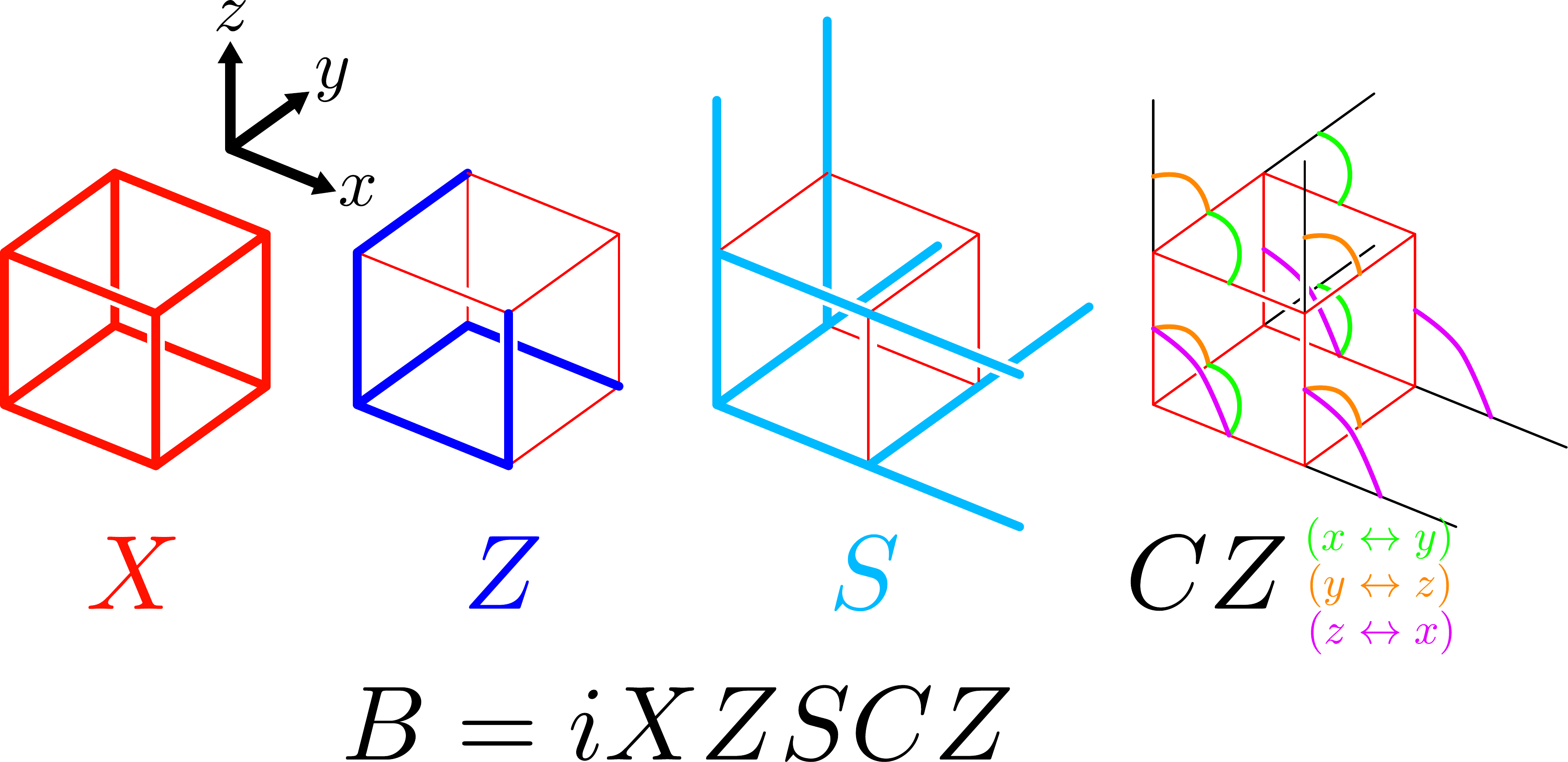}
    \caption{
    The commuting (but non-projector) Hamiltonian describing the fracton dual of the $\mathbb{Z}_2$ strong model is shown.
    Qubit degrees of freedom live on the links.
    The Hamiltonian is a sum over all cubes $c$ of the term $B_c$, shown, which consists of Pauli $X$, $Z$, $S=\mathrm{diag}(1,i)$, and $CZ_{12} = (-1)^{(Z_1-1)(Z_2-1)/4}$ operators between qubits on links of different orientations as shown by the colored lines.
    In addition, the Hamiltonian also has the usual cross terms $A_v^{\mu\nu}$ from the X-Cube which is the product of four $Z$ operators lying in the $\mu\nu$ plane touching a vertex $v$.
    Thus, $H=-\sum_c B_c - \sum_{v}(A_v^{xy} + A_v^{yz} + A_v^{zx})$.
    }\label{fig:strong}
\end{figure}

Conversely, strong 3-foliated SSPTs are dual to fracton models that cannot be realized through such a construction.
This correspondence sheds light on the $F_1$ strong invariant --- in $p$-string condensation, lineon crossing statistics are inherited from the self-braiding statistics of fluxes in the 1-foliated gauge theories, and are therefore the square of a flux exchange statistic, i.e. a multiple of $e^{4\pi \mathrm{i}/N^2}$ for $G=\mathbb{Z}_{N}$ with $N$ even. 
In a strong phase, $F_1=1$ implies that this statistic is offset by $e^{2\pi \mathrm{i}/N^2}$.
The fracton dual of the Type 1 strong $G=\mathbb{Z}_2$ model~\cite{supp} is an example of a novel such fracton order
in which lineons satisfying a triple fusion rule have $\pm i$ mutual crossing statistic, and therefore cannot be realized via $p$-string condensation.
A Hamiltonian realizing this phase is shown in Fig~\ref{fig:strong}.

One can also consider the fracton duals of 2-foliated SSPTs, which are novel `twisted' versions of the 2-foliated lineon-planon model introduced in Ref.~\onlinecite{foliated6}.
Furthermore, the X-cube model may be ungauged in two different ways, by regarding either the fracton sector or the lineon sector as gauge charge. 
The former procedure results in a paramagnet with $G$-valued degrees of freedom transforming under all 3 sets of planar symmetries as before, whereas the latter yields a model with two $G$-valued degrees of freedom per site, the first transforming under $xy$ and $yz$ planar symmetries, and the second under $yz$ and $zx$ planar symmetries. 
 The classification of the latter system is given by $(C_\textrm{2-fol})^2$.
Thus, both Type 1 and Type 2 strong SSPTs, as well as arbitrary weak SSPTs, may be constructed.
Their fracton duals are novel variants of the X-cube model whose fracton dipoles exhibit non-trivial braiding and exchange statistics.

A Type 2 strong SSPT can also be diagnosed through the statistical phases of quasiparticles of the gauged dual.  
Although fractons are immobile particles, we may still define a braiding statistic between two fractons by regarding a single fracton as a semi-infinite stack of fracton dipoles mobile in the $xy$ plane.
Consider a $G=\mathbb{Z}_N\times\mathbb{Z}_N$ model which has two flavors of fractons.
Then, let $e^{\mathrm{i}\theta_{ab}}$ be the statistical phase obtained by braiding two such fractons of flavors $a$ and $b$ on plane $z_0$, where the first argument is a semi-infinite stack in the $z\rightarrow\infty$ direction, and the second argument in the $z\rightarrow-\infty$ direction.
The Type 2 strong invariant is then obtained by $e^{\mathrm{i} F_2 / N} = e^{\mathrm{i}N(\theta_{ab} - \theta_{ba})}$.
This makes it clear why this strong phase with $F_2\neq 0$ cannot be obtained by binding 2D anyons to the fractons, since braiding of 2D anyons is manifestly symmetric with respect to its two arguments.

\emph{Conclusions---}
We have formulated a classification of strong 3D planar SSPTs.
Each phase falls into one of a finite set of equivalence classes modulo stacking with 2D phases, which we have fully enumerated.
For 1-foliated systems, all SSPT phases are weak.
For 2-foliated systems, there are two mechanisms by which a phase may be strong, characterized by Type 1 and Type 2 invariants. 
For 3-foliated systems, only Type 1 strong phases exist.
Under a generalized gauge duality, our classification has a natural interpretation in terms of $p$-string condensation~\cite{coupledlayer}, and we have explicitly constructed strong SSPT models which are dual to fracton phases that cannot be realized via this mechanism.
The fractional quasiparticles in these strong phases thus have novel statistical interactions which cannot be interpreted as the statistics of 2D anyonic bound states.

There are various natural extensions of our work.
A relevant and open question regards the structure of entanglement in strong SSPT phases~\cite{williamsonspee,schmitz,stephenspee,haahspee,bifurcating}.
Another is the addition of non-trivial type-III cocycles, which leads to non-abelian fracton topological orders.
Finally, it would be interesting to study the foliation structure of the fracton duals.

\begin{acknowledgments}
T.D. thanks Fiona Burnell, Dominic Williamson, Abhinav Prem, and Shivaji Sondhi for many helpful discussions, especially in the early parts of this work. 
W.S. thanks Xie Chen and Sagar Vijay for helpful discussions.
T.D. acknowledges support from the Charlotte Elizabeth Procter Fellowship at Princeton University. 
W.S. is supported by the National Science Foundation under award number DMR-1654340 and the Institute for Quantum Information and Matter at Caltech, and by the Simons collaboration on ``Ultra-Quantum Matter''.
J.W. was supported by
NSF Grant PHY-1606531 and Institute for Advanced Study.  This work is also supported by 
NSF Grant DMS-1607871 ``Analysis, Geometry and Mathematical Physics'' 
and the Center for Mathematical Sciences and Applications at Harvard University.
\end{acknowledgments}

\appendix
\title{
Supplementary Material for ``Strong planar subsystem symmetry-protected topological phases  and their dual fracton orders''
}
\author{Trithep Devakul}
\affiliation{Department of Physics, Princeton University, Princeton, NJ 08540, USA}
\author{Wilbur Shirley}
\affiliation{Department of Physics and Institute for Quantum Information and Matter, California Institute of Technology, Pasadena, California 91125, USA}
\author{Juven Wang}
\affiliation{Center of Mathematical Sciences and Applications, Harvard University,  Cambridge, MA 02138, USA}
\affiliation{School of Natural Sciences, Institute for Advanced Study,  Einstein Drive, Princeton, NJ 08540, USA}
\date{\today}
\maketitle

\section{Review of 2D SPTs}\label{app:2dreview}
In this Section, we review the group cohomological classification of SPTs in $2$D, as well as some additional aspects which will prove useful for our arguments related to the SSPT.
These include the interpretation of SPT phases as an anomalous action of the symmetries on the edge, and the connection to the braiding and exchange statistics of quasiparticle excitations in the dual gauge theories.
\subsection{Group cohomological classification of 2D SPTs}

In the presence of symmetry, the unique ground states of two gapped Hamiltonians belong to the the same phase if they can be transformed into each other via a symmetric local unitary (SLU) transformation.\cite{Xie}
That is, a finite depth quantum circuit in which each gate commutes with the symmetry operation.
A state describes a non-trivial $2$D SPT phase if it cannot be connected to the trivial product state via an SLU, but can be trivialized if the symmetry restriction is removed.
Two dimensional bosonic SPTs with on-site symmetry $G$, under this phase equivalence relation, are known~\cite{groupcohomology} to be classified according to the third cohomology group $H^3[G,U(1)]$.
For the finite abelian group $G=\prod_i \mathbb{Z}_{N_i}$, this can be written out explicitly as
\begin{equation}
H^3[G,U(1)] = \prod_i \mathbb{Z}_{N_i} \prod_{i<j} \mathbb{Z}_{\gcd(N_i,N_j)}\prod_{i<j<k} \mathbb{Z}_{\gcd(N_i,N_j,N_k)}
\label{eq:h3class}
\end{equation}
where $\gcd$ denotes the greatest common denominator.
The three factors are commonly referred to as type-I, type-II, and type-III cocycles.
Type-III cocycles correspond to a gauge dual with non-abelian quasiparticle excitations; as our focus is on SSPTs with abelian fracton duals, we will discuss only on type-I and II cocycles.

\subsubsection{The Else-Nayak procedure}\label{app:elsenayak}
Let us derive the group cohomological classification via a series of dimensional reduction procedures, introduced by Else and Nayak.~\cite{elsenayak} which will prove useful in our discussion of SSPTs.
Although the original procedure observed a system with a physical edge, here we prefer to deal with a ``virtual'' edge, meaning: the full system has no edges, but we will consider applying the symmetry only to a finite region $M$ of the system.  
At the edges of $M$, this symmetry will act non-trivially as if at a physical edge.
The advantage of this approach is that it removes any ambiguity related to choice of how the model is defined at the physical edges (and will be useful in the case of SSPTs).

Let $\ket{\psi}$ be the unique gapped ground state of our Hamiltonian $H$ with on-site symmetry group $G$, and $S(g)$ be the symmetry operation realizing the symmetry element $g\in G$.  
We have that $[H,S(g)]=0$ and, without loss of generality, take the ground state to be uncharged under the symmetry $S(g)\ket{\psi} = \ket{\psi}$.
Now, let $S_M(g)$ be the symmetry operation $S(g)$, but restricted to a region $M$.
$S_M(g)$ acting on the ground state will no longer leave it invariant, but will create some excitation along the boundary of this region, $\partial M$.
Since $\ket{\psi}$ is the unique ground state of a gapped Hamiltonian, this excitation may always be locally annihilated by some symmetric unitary transformation $U_{\partial M}(g)^\dagger$, which only has support near $\partial M$.
That is,
\begin{equation}
S_M(g) \ket{\psi} = U_{\partial M}(g) \ket{\psi}
\end{equation}
It is straightforward to show that the matrices $U_{\partial M}(g)$ form a twisted representation of $G$, satisfying
\begin{equation}
\prescript{S_M(g_2)}{}{U_{\partial M}}(g_1) U_{\partial{M}} (g_2) \ket{\psi} = U_{\partial M}(g_1 g_2) \ket{\psi}\label{eq:twistedrep}
\end{equation}
where $\prescript{B}{}A \equiv B A B^{\dagger}$ denotes conjugation of $A$ by $B$, and that they must commute with any global symmetry operation, $[U_{\partial M}(g),S(g^\prime)]=0$.

We now perform a further restriction: from $\partial M$ down to a segment $C$, $U_C(g)$.
This is always possible.
$U_C(g)$ need only satisfy Eq~\ref{eq:twistedrep} up to some unitary operator $V_{\partial C}(g_1,g_2)$ at the two endpoints of $C$,
\begin{equation}
\prescript{S_M(g_2)}{}{U_{C}}(g_1) U_{C} (g_2) \ket{\psi} = V_{\partial C}(g_1,g_2)U_{C}(g_1 g_2) \ket{\psi}
\end{equation}

\begin{figure}
\includegraphics[width=0.45\textwidth]{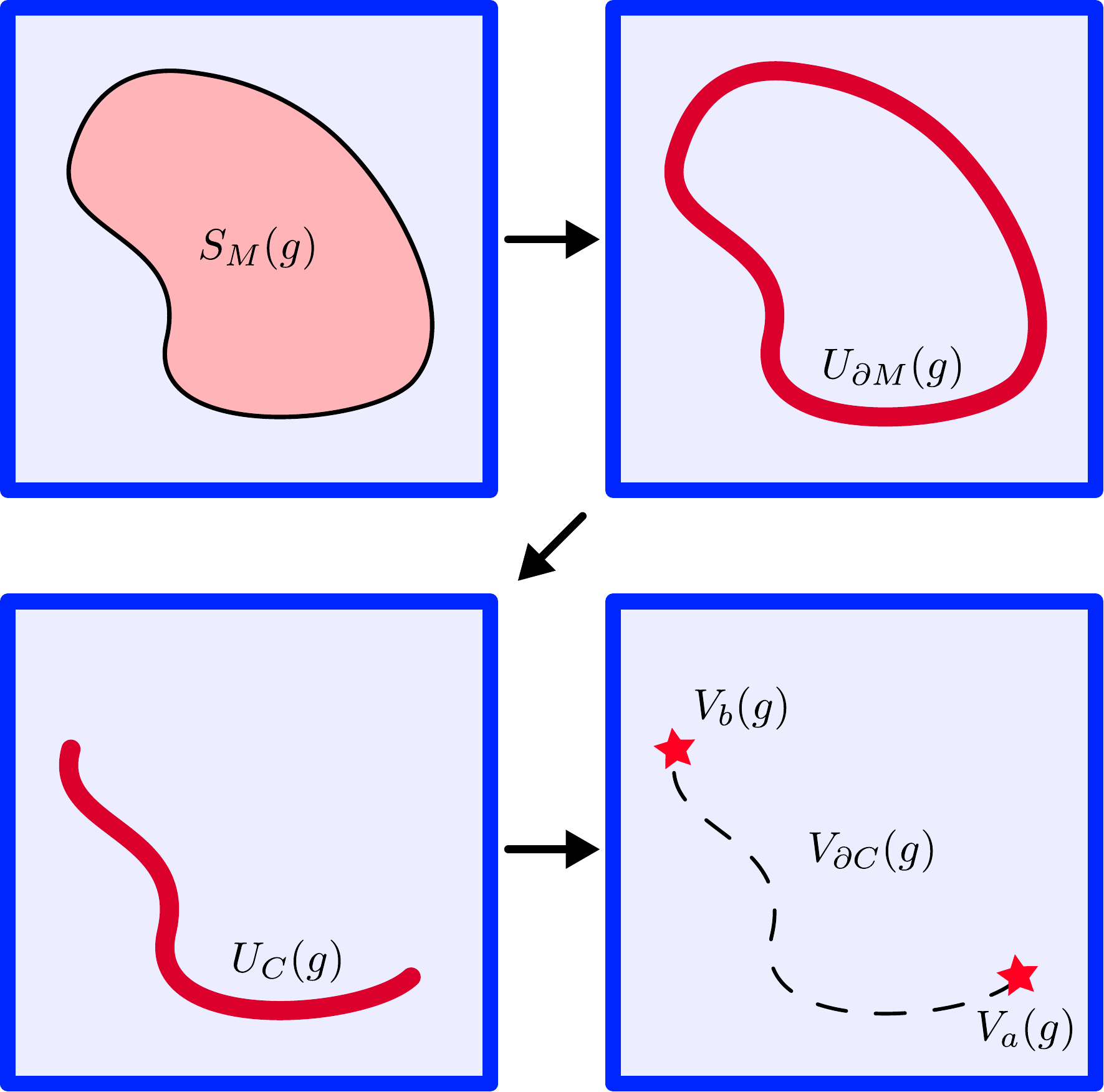}
\caption{
The dimensional reduction procedure in the Else-Nayak procedure.
We start with a truncated global symmetry operator, $S_M(g)$.
This acts on the ground state as a unitary $U_{\partial M}(g)$ along the edge of $M$.
We further restrict this unitary down to a line segment $C$, $U_C(g)$.
Restricted to $C$, $U_C(g)$ behaves as a representation of $G$ only up to unitaries $V_{\partial C}(g)$ at its endpoints.
Finally, we restrict to a single endpoint $V_a(g)$, where associativity of the representation is only satisfied up to a phase $\omega(g_1,g_2,g_3)$, defining our 3-cocycle.
}
\label{fig:elsenayak}
\end{figure}

By associativity, $V_{\partial C}$ must satisfy
\begin{align}
\begin{split}
\prescript{S_M(g_3)}{} V_{\partial C}(g_1,g_2) V_{\partial C}(g_1 g_2, g_3)=
\\ 
\prescript{\prescript{S_M(g_2 g_3)}{} U_C(g_1)}{} V_{\partial C}(g_2,g_3) V_{\partial C}(g_1, g_2 g_3) 
\end{split}\label{eq:associativity}
\end{align}
when acting on $\ket{\psi}$.
The final restriction is from $\partial C$, which consists of two disjoint regions $a$ and $b$, down to simply $a$: $V_{\partial C}(g) = V_{a}(g) V_{b}(g) \rightarrow V_{a}(g) $.
$V_a(g)$ need only satisfy Eq.~\ref{eq:associativity} up to a $U(1)$ phase factor, which can be cancelled out by the contribution from $V_b(g)$.
\begin{align}
\begin{split}
\prescript{S_M(g_3)}{} V_{a}(g_1,g_2) V_{a}(g_1 g_2, g_3)=
\\  
\omega(g_1,g_2,g_3)\prescript{\prescript{S_M(g_2 g_3)}{} U_C(g_1)}{} V_{a}(g_2,g_3) V_{a}(g_1, g_2 g_3) 
\end{split}\label{eq:associativity2}
\end{align}
where $\omega : G^3 \rightarrow U(1)$.
This entire dimensional reduction process is shown in Figure~\ref{fig:elsenayak}.

One can further show that $\omega(g_1,g_2,g_3)$ satisfies the 3-cocycle condition~\cite{elsenayak}
\begin{equation}
1=\frac{\omega(g_1,g_2,g_3)\omega(g_1,g_2 g_3, g_4)\omega(g_2,g_3,g_4)}{\omega(g_1 g_2, g_3, g_4) \omega(g_1,g_2,g_3 g_4)}
\label{eq:cocycle}
\end{equation}
and since $V_a(g_1,g_2)$ is only defined up to a phase factor $\beta(g_1,g_2)$, we must identify
\begin{equation}
\omega(g_1,g_2,g_3) \sim b(g_1,g_2,g_3)\omega(g_1,g_2,g_3) 
\label{eq:cobound}
\end{equation}
where 
\begin{equation}
b(g_1,g_2,g_3) = \frac{\beta(g_1,g_2)\beta(g_1 g_2, g_3)}{\beta(g_2, g_3) \beta(g_1, g_2 g_3)}
\end{equation}
is called a coboundary.
The classification of functions satisfying Eq.~\ref{eq:cocycle}, modulo transformations Eq.~\ref{eq:cobound}, is
exactly the definition of the third cohomology group $H^3[G,U(1)]$.
The class of $\omega$ is the element of $H^3[G,U(1)]$ to which it corresponds.

\subsubsection{Invariant combinations in $H^3$}
Suppose we have followed the Else-Nayak procedure on a system and obtained the cocycle function $\omega(g_1,g_2,g_3)$.  
How do we identify which class in Eq.~\ref{eq:h3class} it belongs to?  
One way to do so is to identify combinations of $\omega$ which are invariant under the transformation Eq.~\ref{eq:cobound}, whose value can tell us about the class.

For simplicity, we focus first on $G=(\mathbb{Z}_N)^{M}$.
Let us first write down an explicit form~\cite{wanggc1,wanggc2} for $\omega$,
\begin{align}
\begin{split}
\omega(g_1, g_2, g_3) =
\exp \left\{\sum_{i \leq j} \frac{2\pi i p^{ij}}{N^2} g_1^{i} (g_2^j + g_3^j - [g_2^j+g_3^j])\right\}
\end{split}\label{eq:explicitcocycle}
\end{align}
where $g^i$ is an integer modulo $N$ denoting the component of $g$ in the $i$th $\mathbb{Z}_N$ factor, $g=(g^1, g^2, \dots , g^M)$, $[\cdot]$ denotes the interior modulo $N$, and $p^{ij}$ are integers mod $N$.
It is straightforward to confirm that $\omega$ satisfies the 3-cocycle condition.
As we will show, the different choices of $p^{ij}$ for $i\leq j$ correspond to different classes in $H^3[G,U(1)]$.  
From Eq.~\ref{eq:h3class}, $p_I^{i}\equiv p^{ii}$ specify the value of the type-I cocycles and $p_{II}^{ij}\equiv p^{ij}$ specify the type-II cocycles.

Define
\begin{equation}
\Omega(g) = \prod_{n=1}^{N} \omega(g,g^n,g)
\end{equation}
and
\begin{equation}
\Omega_{II}(g,h) = \frac{\Omega(gh)}{\Omega(h)\Omega(h)}
\label{eq:app_omega2}
\end{equation}
both of which one can readily verify are invariant under transformations of the type Eq.~\ref{eq:cobound}.
Given a choice of generators, $G=\langle a_1, \dots, a_M\rangle$,
an explicit calculation shows that 
\begin{equation}
\Omega(a_i) = e^{\frac{2\pi i}{N} p_I^{i}}
\end{equation}
and 
\begin{equation}
\Omega_{II}(a_i,a_j)\equiv \frac{\Omega(a_i a_j)}{\Omega(a_i)\Omega(a_j)} = e^{\frac{2\pi i}{N} p_{II}^{ij}}
\end{equation}
thus correctly identifying the value of the type-I and type-II cocycles.
Thus, if we are given an unknown $\omega$, we may simply compute $\Omega(a_i)$ and $\Omega_{II}(a_i,a_j)$ for all $i$ and $j$ to identify its class.

We may define the symmetric matrix $M_{ij} = p_{II}^{ij}$ and $M_{ii} = 2 p_I^{i}$.
Then, we have
\begin{equation}
\Omega(g) = e^{\frac{\pi i}{N} \vec{g}^T \matr{M} \vec{g}}
\label{eq:app_omega}
\end{equation}
and
\begin{equation}
\Omega_{II}(g,h) = e^{\frac{2\pi i}{N} \vec{g}^T \matr{M} \vec{h}}
\label{eq:app_omega2W}
\end{equation}
 for arbitrary elements $g$ and $h$, where $\vec{g}=(g^1,\dots,g^M)$.

\subsubsection{Group cohomology models}
The group cohomology models are a powerful construction that allows us to explicitly write down models realizing SPT phases corresponding to an arbitrary cocycle~\cite{wanggc1,wanggc2}.
Although these models have an elegant interpretation in terms of a path integral on arbitrary triangulations of space-time, we will use them to simply define Hamiltonian models on a square lattice.

We first define the homogenous cocycle $\nu:G^4\rightarrow U(1)$,
\begin{equation}
\nu(g_1, g_2, g_3, g_4) = \omega(g_1^{-1} g_2, g_2^{-1} g_3, g_3^{-1} g_4)
\end{equation}
which satisfies $\nu(g g_1,g g_2, g g_3, g g_4) = \nu(g_1,g_2,g_3,g_4)$.
In terms of $\nu$, the cocycle condition (Eq.~\ref{eq:cocycle}) is
\begin{equation}
1 = \frac{\nu(g_1,g_2,g_3,g_4) \nu(g_1,g_2,g_4,g_5) \nu(g_2,g_3,g_4,g_5)}{\nu(g_1,g_2,g_3,g_5)\nu(g_1,g_3,g_4,g_5)}
\label{eq:nucocycle}
\end{equation}
We will use $\nu$ to define our ground state wavefunction.

Take $G$-valued degrees of freedom on each site $\ve{r}$, $\ket{g_\ve{r}}$.  
The ground state of our model $\ket{\psi}$ is an equal amplitude sum of all possible configurations
\begin{equation}
\ket{\psi} = \sum_{\{g_\ve{r}\}} f(\{g_\ve{r}\}) \ket{\{g_\ve{r}\}}
\end{equation}
where $f(\{g_\ve{r}\})$ is a $U(1)$ phase for each configuration.
The group cohomology model is defined by the choice
\begin{equation}
f(\{g_\ve{r}\}) = \prod_{\ve{r}} \frac{\nu(g_\ve{r}, g_{\ve{r}+\ve{x}}, g_{\ve{r}+\ve{x}+\ve{y}}, g^*)}{\nu(g_\ve{r}, g_{\ve{r}+\ve{y}}, g_{\ve{r}+\ve{x}+\ve{y}}, g^*)}
\equiv \prod_\ve{r} f_\ve{r}(\{g_\ve{r}\})
\label{eq:groupcohomodel}
\end{equation}
where $\ve{x},\ve{y}$ are the two unit vectors, $g^*\in G$ is an arbitrary element which we can simply take to be the identity $g^*=1$, and we have defined a phase contribution $f_\ve{r}$ for each plaquette.
This arises from a triangulation of each square plaquette into two triangles, each of which contribute a phase; those interested in the details of the construction are directed to Ref~\onlinecite{wanggc1}.


Performing the Else-Nayak procedure outlined in Section~\ref{app:elsenayak} on this ground state results in exactly the cocycle $\omega$ used to construct the state, up to a coboundary (Eq.~\ref{eq:cobound}).

To obtain a gapped local Hamiltonian realizing this state as its ground state, we simply consider a set of local ergodic transitions $\langle\{g_\ve{r}\}\rightarrow\{g_\ve{r}^\prime\}\rangle$, multiplied by an appropriate phase factor,
\begin{equation}
H = -\sum_{\langle \{g_\ve{r}^\prime\}\rightarrow \{g_\ve{r}\}\rangle} \frac{f(\{g_\ve{r}^\prime\})}{f(\{g_\ve{r}\})} \ket{\{g_\ve{r}^\prime\}}\bra{\{g_\ve{r}\}}
\label{eq:gcham}
\end{equation}
which by construction has $\ket{\psi}$ as its unique ground state.
We can simply choose $\{g_\ve{r}^\prime\}$ to differ from $\{g_\ve{r}\}$ by the action of a generator $a_i$ of $G$ on a single site $\ve{r}$.
The Hamiltonian will then be a sum of mutually commuting terms consisting of a ``flip'' operator $\ket{a_i g_\ve{r}}\bra{g_\ve{r}}$ on each site, multiplied by an appropriate phase factor depending on the state $\{g_\ve{r}\}$ near that site.

\subsubsection{Gauge duality}
The group cohomological classification of an SPT has an elegant interpretation in terms of braiding statistics of its gauge dual.~\cite{levingu}
We will briefly outline the gauging process (as applied to the group cohomology models), and discuss the relevant statistical processes.

Consider the group cohomology SPT model on a square lattice given by Eq.~\ref{eq:gcham}.
To gauge the global symmetry, we define gauge degrees of freedom $g_{\ve{r},\ve{r}^\prime}=g_{\ve{r}^\prime,\ve{r}}^{-1}$ for each nearest neighbor pair $(\ve{r},\ve{r}^\prime)$, and enforce a Gauss's law constraint at each vertex $\ve{r}$ which involves the matter degree of freedom $g_{\ve{r}}$ and the adjacent gauge degrees of freedom $g_{\ve{r},\ve{r}'}$. Then, we minimally couple the symmetric Hamiltonian to the gauge degrees of freedom by replacing the operators $g_{\ve{r}^\prime}g_{\ve{r}}^{-1}$ with $g_{\ve{r}^\prime}g_{\ve{r},\ve{r}^\prime}g_{\ve{r}}^{-1}$ throughout.
In addition, we energetically enforce the zero-flux constraint $g_{\ve{r}_1 \ve{r}_2} g_{\ve{r}_2 \ve{r}_3} g_{\ve{r}_3 \ve{r}_4} g_{\ve{r}_4 \ve{r}_1}=1$
for the square plaquette with corners $\ve{r}_{1\dots 4}$ (labeled going clockwise or counterclockwise), by adding an appropriate projection term to the Hamiltonian.

The resulting model describes a topologically ordered phase, with characteristic properties such as a topological ground state degeneracy on a torus and quasiparticle excitations with anyonic braiding statistics.
There are two types of excitations: gauge charge, denoted by $e_g$, and gauge flux, denoted by $m_g$, for each $g\in G$.
The former are created by gauged versions of operators of the form 
\begin{equation}
Z_g^\dagger (\ve{r}_1) Z_g(\ve{r}_2) = \sum_{\{g_\ve{r}\}}e^{\frac{2\pi i}{N}(g_{\ve{r}_2}^i - g_{\ve{r}_1}^i)}\ket{\{g_\ve{r}\}}\bra{\{g_\ve{r}\}}
\end{equation}
which creates a charge-anticharge pair, $e_g$ and $e_g^{-1}$, at positions $\ve{r}_2$ and $\ve{r}_1$.
To create gauge flux excitations, instead consider the gauged version of the operator
\begin{equation}
L(g) \ket{\psi}\equiv U_{\partial M}^{\dagger}(g) S_M(g) \ket{\psi} = \ket{\psi}
\end{equation}
where $S_M(g)$ is a symmetry operator restricted to a region $M$ and $U_{\partial M}(g)$ is the action on the boundary $\partial M$, as in the dimensional reduction procedure of Section~\ref{app:elsenayak}.
The gauged version of $S_M(g)$ only flips $g_{\ve{r}, \ve{r}^\prime}$ near at the boundary, and so the gauged $L(g)$ operator has support only on $\partial M$.  
Now, if we further restrict $L(g)\rightarrow L_C(g)$ to an open segment $C$, $L_C(g)$ creates two quasiparticle excitations at the two endpoints, which we identify as the gauge flux-antiflux pair $m_g$ and $m_g^{-1}$.
Note that there is an ambiguity in defining the gauged version of $L(g)$, which may result in a different definition of the gauge flux excitation, $m_g\sim m_g e_{g^\prime}$.
Thus, gauge fluxes are only well defined modulo attachment of charges.

The group cohomological classification of the ungauged SPT manifests in the self and mutual statistics of gauge fluxes in the gauged theory.
Let $a_i$ be the generator of the $i$th factor of $\mathbb{Z}_N$ in $G$, and $e_i$ and $m_i$ be its gauge charge and flux excitations.
For two identical excitations, we can define an exchange phase via a process in which their two positions are exchanged.
For two different excitations, we may instead define the full braiding phase, which is accumulated when one particle encircles another.
In the gauge theory, $e_i$ all have trivial exchange and only braid non-trivially with its own gauge flux $m_i$.
Meanwhile, the gauge flux $m_i$ has an exchange statistic $e^{\frac{2\pi i p_I^{i}}{N^2}}$ with itself, and a mutual braid $e^{\frac{2\pi i p_{II}^{ij}}{N^2}}$ with $m_j$.
Notice that the exchange and mutual braid of $m_i$ is only well defined modulo $e^{\frac{2\pi i}{N}}$, since $m_i$ is only well defined modulo charge attachment.
For a general gauge flux $m_g$, its exchange phase is given by an $N$th root of $\Omega(g)$, which can be straightforwardly calculated from the $\matr{M}$ matrix (Eq.~\ref{eq:app_omega}).

For those familiar with the $K$ matrix formulation of Abelian topological orders, we note that the $K$ matrix characterizing the gauged theory is given by 
\begin{equation}
\matr{K} = 
\begin{bmatrix}
-\matr{M} & N \matr{1}\\
N \matr{1} & 0
\end{bmatrix}
\end{equation}
such that 
\begin{equation}
\matr{K}^{-1} = 
\begin{bmatrix}
0 & \frac{1}{N}\matr{1}\\
\frac{1}{N}\matr{1} & \frac{1}{N^2} \matr{M}
\end{bmatrix}
\end{equation}
where the indices represent quasiparticle excitations ordered as $\{e_1, e_2, \dots, m_1, m_2, \dots \}$.
The exchange statistic of a quasiparticle $\vec{l}$ written in this basis is given by $e^{\pi i \vec{l}^T \matr{K^{-1}} \vec{l}}$, and the mutual braiding statistic between $\vec{l}_1$ and $\vec{l}_2$ is $e^{2\pi i \vec{l}_1^T \matr{K}^{-1} \vec{l}_2}$.

\section{The symmetry action in SSPTs}\label{app:symac}
Let us briefly discuss how symmetries may act anomalously on the edges in an SSPT.
Again, let us consider only virtual edges, as in our earlier 2D discussion.

We first discuss the case for 3-foliated phases.
Let us take a cubic subregion $M$, and consider applying symmetry operations restricted to this subregion.  
An $xy$ planar symmetry restricted to this region, $S^{xy}_M(z;g)$, will act on the ground state as some unitary operation along the boundary near the plane $z$,
\begin{equation}
S^{(xy)}_M(z;g) \ket{\psi} = U_{\partial M}(z;g)\ket{\psi}
\end{equation}
exactly as for the 2D SPT earlier.  
The same is true for $yz$ or $zx$ planar symmetries.

\begin{figure}
\includegraphics[width=0.45\textwidth]{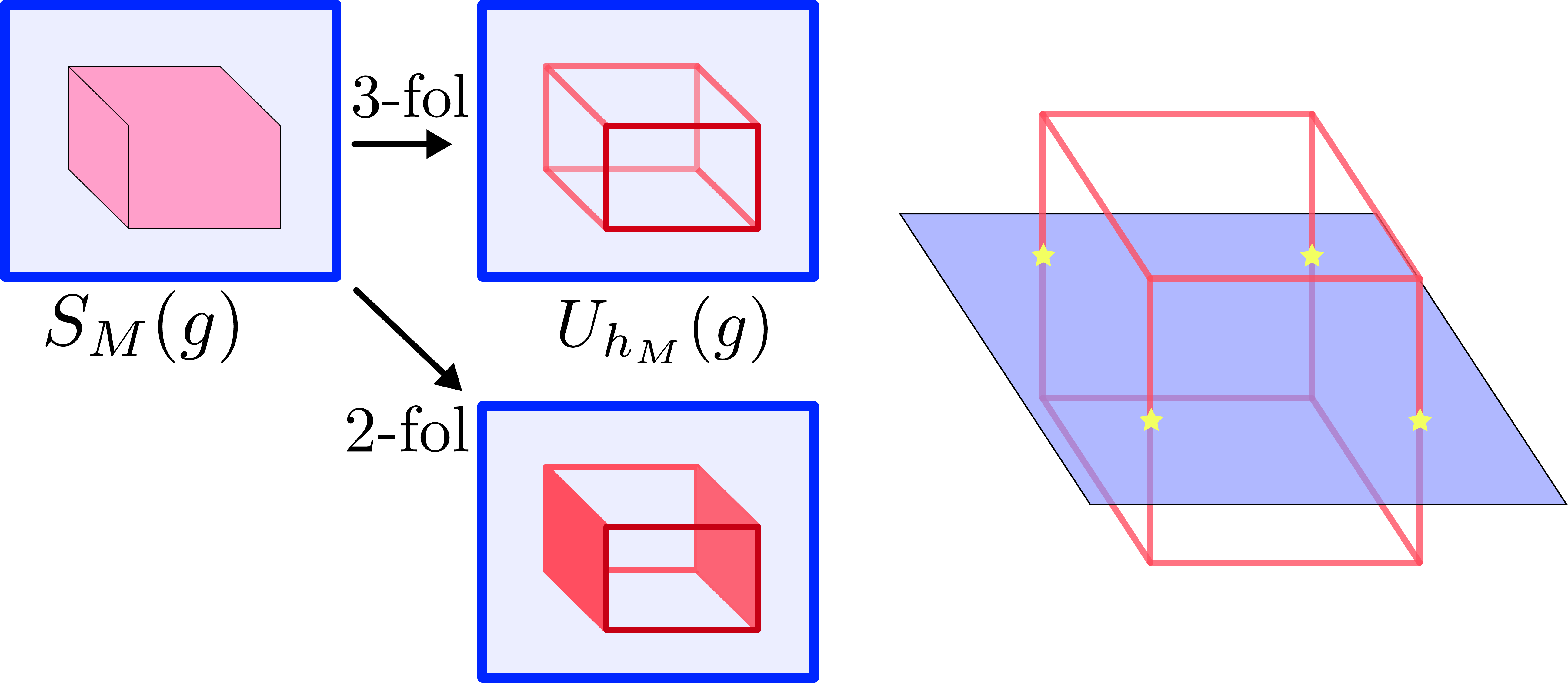}
\caption{
(Left) The action of a symmetry restricted to a large cube, $S_M(g)$, acts on the ground state as a unitary supported along the hinges $U_{h_M}(g)$ in a 3-foliated model.
In the 2-foliated model with only $xy$ and $yz$ planar symmetries, it may act non-trivially along the $zx$ face of $M$ as well.
(Right)
A phase may be non-trivial if the hinges of the cube operator contain non-trivial charge under a planar symmetry.
This type of non-triviality does not show up in its $H^3$ classification, but can be generated by stacking 2D linear SSPTs.
}\label{fig:symac}
\end{figure}

Now, let us instead consider applying the global symmetry restricted to this subsystem, which we call simply $S_M(g)$.
Along the $xy$ faces of the cube $M$, $S_M(g)$ looks like a symmetry operation $S^{xy}_M(z;g)$, and similarly along the $yz$ and $zx$ faces.
Thus, $S_M(g)$ acting on the ground state cannot act non-trivially along the edges.
The only place where $S_M(g)$ does not look like a symmetry operator is along the hinges of $M$, which we denote by $h_M$.
Thus,
\begin{equation}
S_M(g)\ket{\psi} = U_{h_M}(g)\ket{\psi}
\end{equation}
acts as some unitary operator along the hinges of $M$.
For 2-foliated phases, we may apply the same argument, except that only two of the planar symmetries exist.  
Suppose we only have $xy$ and $yz$ planar symmetries.  
Then, $S_M(g)$ acting on the ground state may act non-trivially along the hinges and the $zx$ face of $M$.  
This is shown in Fig.~\ref{fig:symac} for the 3 and 2-foliated cases.

Knowing the way the symmetry acts along the edges of $M$ is sufficient to obtain the $H^3[G^L,U(1)]$ classification of the phase.  
For example, one can readily apply the Else-Nayak procedure detailed earlier in order to extract the cocycle function $\omega$.  
However, there is more information contained in $U_{h_M}(g)$ that is missed in this process.
We know that $U_{h_M}(g)$ must commute with all untruncated symmetries, such as $S^{(xy)}(z;g)$, when acting on the ground state.  
That is, $U_{h_M}(g)$ has to be overall charge neutral under all symmetries.
Consider the symmetry $S^{(xy)}(z;g^\prime)$ which intersects with four hinges in $U_{h_M}(g)$, as in Fig.~\ref{fig:symac}.  
The four intersection locations are spatially separated, thus one can sensibly define a charge on each of the four hinges, which do not have to be trivial.  
That is, $S^{xy}(z;g^\prime)$ commuting with the first hinge may result in a phase $e^{i \phi_{1}}$, the second $e^{i\phi_2}$, and so forth, which is fine as long as $e^{i (\phi_1+\phi_2+\phi_3+\phi_4)}=1$.
These charges pinned to the hinges cannot be removed by a symmetric local unitary transformation, and are therefore a sign of a non-trivial phase.
Such charges arise due to the existence of 2D linear SSPTs: 2D phases with SSPT order protected by line-like subsystem symmetries.~\cite{dwy}
In the following section we will construct an explicit example of this.

\section{Non-trivial SSPT phases with trivial $H^3$}\label{app:semionic}
In this section, we highlight a mechanism by which an SSPT phase may be non-trivial, despite appearing trivial in our $H^3[G^L,U(1)]$ picture along all planar directions (but still a weak phase overall).
Let us begin with an example: the so-called semionic X-cube model (See Eq.~14 and Fig.~9 in Ref.~\onlinecite{foliated4}, also Ref.~\onlinecite{coupledlayer}).
This is a $G=\mathbb{Z}_2$ fracton gauge theory in which lineons satisfying a triple fusion rule have a $-1$ braiding statistic with one another.  
There is a qubit degree of freedom on each site, which are acted on by Pauli matrices $X$ and $Z$.
The ungauged model is given by the Hamiltonian
\begin{equation}
H_\text{sem} = -\sum_\ve{r} X_\ve{r} C_\ve{r} C_{\ve{r}-\ve{x}-\ve{y}-\ve{z}}
\end{equation}
where
\begin{equation}
C_\ve{r} = \prod_{a=0}^{1}\prod_{b=0}^{1}\prod_{c=0}^{1} Z_{\ve{r}+a\ve{x}+b\ve{y}+c\ve{z}}
\end{equation}
is a product of 8 $Z$s on the corners of a cube.

We immediately notice that this model has a trivial $H^3[G^L,U(1)]$, which one can confirm by simply noting that a planar symmetry actually acts trivially along the edges (except at a corner), and so cannot produce anything non-trivial under the Else-Nayak procedure.
This is due to the presence of higher symmetry: this model actually is symmetric under \emph{line-like} symmetries.
For example, a product of $X$s along the $x$ direction,
\begin{equation}
S^{x}(y,z;g) = \prod_{x} X_{\ve{r}=(x,y,z)}
\end{equation}
commutes with $H_\text{sem}$.
Acting on the ground state, the global symmetry operator truncated to a cube $S_M(g)$ acts by creating charges localized at its corners (as it must be for a 3D system with line-like subsystem symmetries).  
It is this pattern of charges which leads to the non-trivial lineon braiding phase.

Notice that there is no contradiction between the system having a trivial $H^3[G^L,U(1)]$ classification and lineons having a $-1$ braiding statistic.
This is due to the fact that the fundamental braiding process which $H^3[G^L,U(1)]$ cares about is between lineon dipoles.
Braiding two lineons in a plane $z_0-1/2$ is like braiding a stack of lineon dipoles on planes $z<z_0$ with another stack $z\geq z_0$.
However, the braiding phase in a $\mathbb{Z}_2$ theory is only defined modulo $\pm 1$, and so a braiding phase of $-1$ is the same as trivial from this perspective.

It is straightforward to show that the model described by $H_\text{sem}$ is weak.  
One can write the Hamiltonian as
\begin{equation}
H_\text{sem} = -\sum_\ve{r} U X_\ve{r} U^\dagger
\end{equation}
where $U$ is a local unitary circuit consisting of $CZ$ gates.  
The ground state is then
\begin{equation}
\ket{\psi} = U \ket{\psi_0}
\end{equation}
where $\ket{\psi_0}$ is the trivial paramagnetic phase.
It is possible to write $U$ as
\begin{equation}
U = \prod_{z} U_z
\end{equation}
where $U_z$ acts only between layers $z$ and $z+1$, and $U_z$ commutes with all planar symmetries (and each other)
This is exactly the form of a planar-symmetric local unitary circuit (just a higher dimensional version of the linearly-symmetric local unitary circuit defined in  Ref~\onlinecite{dwy}).
Thus, $U_z\ket{\psi_0}$, where $\ket{\psi_0}$ is the trivial paramagnetic state $X_\ve{r}\ket{\psi_0}=\ket{\psi_0}$, describes a 2D phase on planes $z$, $z+1$ (which is actually a 2D linear SSPT), and the ground state of $H_\text{sem}$ is simply the stack of these.

One of the consequences of the fact that $H_\text{sem}$ describes a weak phase is that there is no obstruction to constructing an SSPT phase which is described by $H_\text{sem}$ for $z\ll 0$ and is completely trivial $H_\text{triv}$ for $z\gg 0$.
For example, we can define the Hamiltonian
\begin{align}
\begin{split}
H_\text{half} =& -\sum_{\ve{r}: r_z < 0}X_\ve{r} C_\ve{r} C_{\ve{r}-\ve{x}-\ve{y}-\ve{z}}
 - \sum_{\ve{r}:r_z>0}X_\ve{r}
\\
 &- \sum_{\ve{r}:r_z=0}X_\ve{r} Z_\ve{r}Z_{\ve{r}+\ve{x}} Z_{\ve{r}+\ve{y}} Z_{\ve{r}+\ve{x}+\ve{y}} C_{\ve{r}-\ve{x}-\ve{y}-\ve{z}}
\end{split}
\end{align}
which is composed of commuting terms.
If we look at the action of $S_M(g)$  where $M$ is a large cube crossing $z=0$, one finds that there is a single $Z$ pinned to each hinge at $z=0$,
exactly as discussed in Section~\ref{app:symac}.
Indeed, by stacking 2D SSPTs, it is possible to realize phases with various choices of allowable charges pinned on each hinge.

Finally, we note that we do not have a proof that \emph{all} phases with a trivial $H^3[G^L,U(1)]$ classification are weak.
There may also exist other mechanisms by which a phase may be non-trivial.
However, we are not aware of any counterexamples.

\section{Mobility of single lineons}\label{app:deconflineon}
Here, we show that in the fracton dual of a 3-foliated phase, a ``single lineon'' need not actually be mobile along lines. 
In a slight abuse of nomenclature, we will still call this excitation the lineon, even though it may not be mobile along a line.

First, let us identify what is commonly referred to as the lineon.
Again, consider the action of the global symmetry truncated to a cube, $S_M(g)$, which acts on the ground state as some unitary along the hinges $h_M$,
\begin{equation}
S_M(g)\ket{\psi} = U_{h_M}(g)\ket{\psi}
\end{equation}
Then, the operator $L_M(g)\equiv S_M(g) U_{h_M}^\dagger (g)$ acts trivially on the ground state.
The gauged version of the operator $L_M(g)$ will define our lineon.

Let us review how the generalized gauging process works for a 3-foliated planar symmetric model (see Refs.~\onlinecite{vijayhaahfugauge,williamsongauge,shirleygauge} for details).
For each $xy$ plaquette, we introduce the plaquette gauge degree of freedom $\tilde{g}_\ve{r}^{(xy)}$ which resides on the plaquette centered at $\ve{r}+\frac{1}{2}\ve{x}+\frac{1}{2}\ve{y}$,
and similarly for $yz$ and $zx$ plaquettes. A Gauss's law constraint is subsequently enforced at each vertex $\ve{r}$ which involves the matter degree of freedom $g_\ve{r}$ and the 12 adjacent gauge degrees of freedom. 
We then minimally couple the subsystem symmetric Hamiltonian to the gauge degrees of freedom by replacing the symmetric coupling terms $g_\ve{r} g_{\ve{r}+\ve{x}}^{-1} g_{\ve{r}+\ve{y}}^{-1} g_{\ve{r}+\ve{x}+\ve{y}}$ with the term $\tilde{g}_\ve{r}^{(xy)}g_\ve{r} g_{\ve{r}+\ve{x}}^{-1} g_{\ve{r}+\ve{y}}^{-1} g_{\ve{r}+\ve{x}+\ve{y}}$ throughout.
Finally, we energetically enforce the zero-flux constraints
\begin{align}
1 = \tilde{g}_\ve{r}^{(zx)}
\tilde{g}_{\ve{r}+\ve{x}}^{(yz)}
\left(\tilde{g}_{\ve{r}+\ve{y}}^{(zx)}
\tilde{g}_{\ve{r}}^{(yz)}\right)^{-1}
\\
1 = \tilde{g}_\ve{r}^{(xy)}
\tilde{g}_{\ve{r}+\ve{y}}^{(zx)}
\left(\tilde{g}_{\ve{r}+\ve{z}}^{(xy)}
\tilde{g}_{\ve{r}}^{(zx)}\right)^{-1}
\\
1 = \tilde{g}_\ve{r}^{(yz)}
\tilde{g}_{\ve{r}+\ve{z}}^{(xy)}
\left(\tilde{g}_{\ve{r}+\ve{x}}^{(yz)}
\tilde{g}_{\ve{r}}^{(xy)}\right)^{-1}
\end{align} 
by adding terms to the Hamiltonian which projects onto this subspace.

In terms of these plaquette variables, the symmetry $S_M(g)$ acts only on the hinges of the cube $M$.
If the gauged version of $U_{h_M}$ can be written such that it only acts along the hinges as well, then the gauged version of $L_M(g)=S_M(g)U_{h_M}^\dagger$ also only acts along the hinges.
In this case, if we truncate $L_M(g)$, we obtain an operator which creates a $g$ lineon excitation at each of its truncated hinges.
A single lineon is guaranteed to be mobile along a line, simply due to the fact that the operator $L_M(g)$ is line-like along the hinges.
While this has been true in virtually all previously studied models, it is not true generally.

Indeed, consider $H_\text{half}$ from Section~\ref{app:semionic}.
In that model, $U_{h_M}(g)$ has a single charge $Z_\ve{r}$ pinned at each place where $h_M$ crossed $z=0$.
However, there is no way to gauge $U_{h_M}(g)$ in such a way as to keep the support of the operator only along the hinges.
This means that one cannot construct a lineon which crosses the $z=0$ plane alone.

From the perspective of the fracton order, we may consider a $z$-moving lineon at $z>0$.  
Now, suppose we naively move this lineon down to $z<0$, crossing the $z=0$ plane. 
What one will find is that, upon crossing, there is a single fracton charge excitation stuck at the $z=0$ plane.  
As it is a single fracton, which is immobile, one cannot simply move it along with the lineon (which would simply amount to a redefinition of the lineon for $z<0$ vs $z>0$).
Thus, a $z$-moving lineon cannot cross the $z=0$ plane without paying an energy penalty in the form of a fracton stuck at $z=0$.

Now, instead of having simply a single plane at $z=0$ where charges are pinned, we can imagine constructing a model in which charges are pinned on every plane, or every other plane, for example.
In this case, a single lineon moving would create fracton excitations as it moved along, which are unable to be annihilated or moved along with the lineon as a redefinition.
A single lineon therefore cannot be moved along a line without creating additional excitations.
However, a pair of lineon anti-lineon on adjacent planes (the gauge flux) is always guaranteed to be a planon.

\section{Proof of constraints}\label{app:constraints}
Here, we prove the two constraints mentioned in the main text for 2 or 3-foliated models.
Let us label by $g_z$ the group element $g$ in the $z$th factor of $G^L$, and $g_\text{glo}=\prod_z g_z$ a global symmetry.
Again, we view the SSPT as a quasi-2D system with the large symmetry group $G^L$.
We will denote the representation of the symmetry $g$ acting on the $z$th plane by simply $S(g_z)$, rather than $S^{xy}(z;g)$ as in the main text.
We take the system to also be symmetric under $yz$-planar symmetries.

Consider a square region $M$ of this quasi-2D system ($M$ would contain all sites $x,y,z$ with $x_1<x<x_2$, $y_1<y<y_2$, and all $z$, for some choice of $x_{1,2},y_{1,2}$).
The key fact is that the global symmetry truncated to $M$, $S_M(g_\text{glo})$, acts trivially along the $yz$ face of $M$ (simply due to the fact that it acts identically to $yz$-planar symmetries).
Thus, $U_{\partial M}(g_\text{glo})$ acts trivially along the $yz$ face.
Now, we may perform the Else-Nayak procedure, further choosing a restriction to an open segment $C$ which ends along the $yz$ face.
Going through the procedure with a trivial $U_{\partial M}(g_\text{glo})$, 
we can always get $\omega(g_\text{glo},h_\text{glo},k_\text{glo})=1$ for arbitrary $g,h,k\in G$ (up to a coboundary).

This leads to our second constraint.
Calculating the invariant $\Omega(g)=\prod_{n=1}^N \omega(g,g^n,g)$ for any global symmetry results in 
a trivial type-I cocycle with itself, $ \Omega(g_\text{glo})=1 $.
In terms of the $\matr{M}$ matrix, 
\begin{equation}
\Omega(g_\text{glo}) = e^{ \pi i \vec{g}_\text{glo}^T \matr{M} \vec{g}_\text{glo} / N} = 1
\end{equation}
for each generator of $G$ is exactly the global constraint from the main text.

Next, consider the type-II cocycle between a global symmetry $h_\text{glo}$ and $g_z$.
This is given by the ratio
\begin{equation}
\Omega_{II}(g_{z},h_\text{glo}) = \Omega( g_zh_\text{glo})/(\Omega(g_z)\Omega(h_\text{glo}))  .
\end{equation}
We can calculate $\Omega(h_\text{glo} g_z)$ using the Else-Nayak procedure, which we wish to show is simply equal to $\Omega(g_z)$.

First, note that if we have $S_{M^\prime}(g)$ defined on some larger $M^\prime$, which has a boundary action $U_{\partial M^\prime}(g)$, we may always use this to construct an edge action for $S_M(g)$ as
\begin{align}
\begin{split}
S_M(g)\ket{\psi} &= S_{M}(g) S_{M^\prime}^\dagger(g) S_{M^\prime}(g)\ket{\psi}\\
&= S_{M}(g) S_{M^\prime}^\dagger(g) U_{\partial M^\prime}(g)\ket{\psi}\\
&\equiv U_{\partial M}(g)\ket{\psi}
\end{split}
\label{eq:symextend}
\end{align}
which acts simply as $S_M(g) S_{M^\prime}^\dagger(g)$ near $\partial M$, and has deferred all the non-triviality over to $\partial M^\prime$.
Now, we may use this construction for $U_{\partial M}(g_z)$ in the Else-Nayak procedure, which, along the $yz$ face, is equivalent to $U_{\partial M}(g_z h_\text{glo})$ (since $U_{\partial M}(g_z)=1$ is trivial along this edge). 
The procedure then continues, and since $U_{\partial M}(g_z h_\text{glo})$ is (by construction) invariant under conjugation by $S_M(h_\text{glo})$, the process proceeds exactly the same regardless of whether we had chosen $g_z h_\text{glo}$ or just $g_z$.  We can therefore always choose to have
\begin{equation}
\omega(g_z h_\text{glo},(g_z h_\text{glo})^n,g_z h_\text{glo}) = 
\omega(g_z ,(g_z )^n,g_z )  
\end{equation}
so that $\Omega(g_z h_\text{glo}) = \Omega(g_z)$, and therefore $\Omega_{II}(g_z,h_\text{glo}) = 1$.
In terms of the $\matr{M}$ matrix,
\begin{equation}
\Omega_{II}(g_z,h_\text{glo}) = e^{2\pi i \vec{h}_z^T \matr{M} \vec{g}_\text{glo} / N} = 1\label{eq:applocalconstraint}
\end{equation}
for $h$, $g$, being generators of $G$, is exactly the local constraint in the main text.

\section{Various proofs for invariants $F_1$ and $F_2$}\label{app:proofs}
\subsection{Independence of direction for $F_1$ and $F_2$, and triviality of $F_2$ in 3-foliated model}
In this section, we prove the claims in the main text that 1) the invariants $F_1$ and $F_2$ must be the same regardless of which direction of planar symmetry we look at, and 2) that $F_2$ must be trivial in a 3-foliated model.

We first introduce some ideas for a regular 2D SPT.
First, let us make the simplifying assumption that $U_{\partial M}(g)$ is a purely diagonal operator.
This is always possible to do in our class of models, where $\ket{\psi}$ is an equal amplitude sum
\begin{equation}
\ket{\psi} = \sum_{\{g_\ve{r}\}}f(\{g_\ve{r}\})\ket{\{g_\ve{r}\}}
\end{equation}
since if $S_M(g)$ sends $\{g_\ve{r}\}\rightarrow\{g_\ve{r}^\prime\}$, then we may simply choose
\begin{equation}
U_{\partial M}(g) = \sum_{\{g_\ve{r}\}} \frac{f(\{g_\ve{r}\})}{f(\{g_\ve{r}^\prime\})} \ket{\{g_\ve{r}^\prime\}}\bra{\{g_\ve{r}^\prime\}}
\end{equation}
which one can verify satisfies $U_{\partial M}^\dagger(g) S_M(g)\ket{\psi}=\ket{\psi}$ and will be only supported along $\partial M$ as $\ket{\psi}$ is symmetric.
Note that although we have made this assumption, the spirit of our argument should remain the same even without it.
In the Else-Nayak procedure, this means that $U_C(g)$ and $V_{\partial C}(g_1,g_2)$ can also be chosen to be purely diagonal, and Eq.~\ref{eq:associativity2} reads
\begin{align}
\begin{split}
\prescript{S_M(g_3)}{}V_a(g_1,g_2)V_a(g_1g_2,g_3)
= \\
\omega(g_1,g_2,g_3)  V_a(g_2,g_3) & V_a(g_1,g_2 g_3)\\
\end{split}\label{eq:diageqn}
\end{align}

To measure $\Omega(g)$,
consider the product 
\begin{equation}
Q_a(g) = \prod_{n=1}^N V_a(g, g^n)\label{eq:Qdef}
\end{equation}
which one can show using Eq.~\ref{eq:diageqn} satisfies
\begin{equation}
\prescript{S_M(g)}{}Q_a(g) = \Omega(g)Q_a(g) 
\label{eq:omegacalc}
\end{equation}
That is, the charge of $Q_a(g)$ under $S_M(g)$ is exactly the type-I invariant $\Omega(g)$.
This procedure has the nice interpretation in the gauged language as measuring (half) the charge of $N$ gauge fluxes $m_g$.  

Next, consider a measurement of $\Omega_{II}(g,h)$.  
One way to do so is by noting that we may use a region $M_1$ for $S_M(g)=S_{M_1}(g)$, but instead consider a much larger region $M_2$ which fully contains $M_1$ for the symmetry $S_{M_2}(h)$, and also define $S_M(g^n h^n) = S_{M_1}(g^n)S_{M_2}(h^n)$
(formally, we would absorb some of the symmetry into $U_{\partial M}(h)$, like in Eq~\ref{eq:symextend}).
Then, using the fact that $U_{\partial M}(g)$ commutes with all full symmetries $S(h)$, and $S_{M_2}(h)\approx S(h)$ when acting on $U_{\partial M_1}(g)$ since $M_2$ is much larger than $M_1$, we have
\begin{equation}
U_{\partial M}(gh) = U_{\partial M_1}(g) U_{\partial M_2}(h)
\end{equation}
Next, one can always choose the truncation to a segment $U_{\partial M_1}(g)\rightarrow U_{C_1}(g)$ in a way that $U_{C_1}(g)$ also commutes with all full symmetries $S(h)$, in which case
\begin{equation}
V_{\partial C} (g h, g^n h^n) = V_{\partial C_1}(g,g^n) V_{\partial C_2}(h,h^n)
\end{equation}
as well.
Using this choice, we have that $Q_a(gh) = Q_{a_1}(g) Q_{a_2}(h)$.

From this, one can readily compute the type-II cocycle
\begin{equation}
\Omega_{II}(g,h) = \frac{\prescript{S(h)}{} Q_a(g)}{Q_a(g)}
\label{eq:omega2calc1}
\end{equation}
And by symmetry, 
\begin{equation}
\Omega_{II}(g,h) = \frac{\prescript{S(g)}{} Q_a(h)}{Q_a(h)}
\label{eq:omega2calc2}
\end{equation}
(note that these expressions are unambiguous since both numerator and denominator are diagonal).
These have the nice interpretation in the gauged picture of measuring the number of charges $e_h$ obtained as a fusion result of $N$ gauge fluxes $m_g$, or vice versa.

Notice that while $Q_a(g)$ carries a charge under $S_M(g)$ and $S(h)$, if we consider the contribution from the other endpoint of $\partial C$, $Q_b(g)$, then one must have that
\begin{align}
\begin{split}
\prescript{S_M(g)}{} (Q_a(g) Q_b(g)) = Q_a(g) Q_b(g)\\
\prescript{S(h)}{} (Q_a(g) Q_b(g)) = Q_a(g) Q_b(g)
\end{split}
\label{eq:zerocharge}
\end{align}
the phase factors cancel out from the two endpoints.
This is simply due to the fact that the phase $\omega$ only appears when isolating $V_{\partial C}(g)$ to a single endpoint.

\begin{figure}
\includegraphics[width=0.5\textwidth]{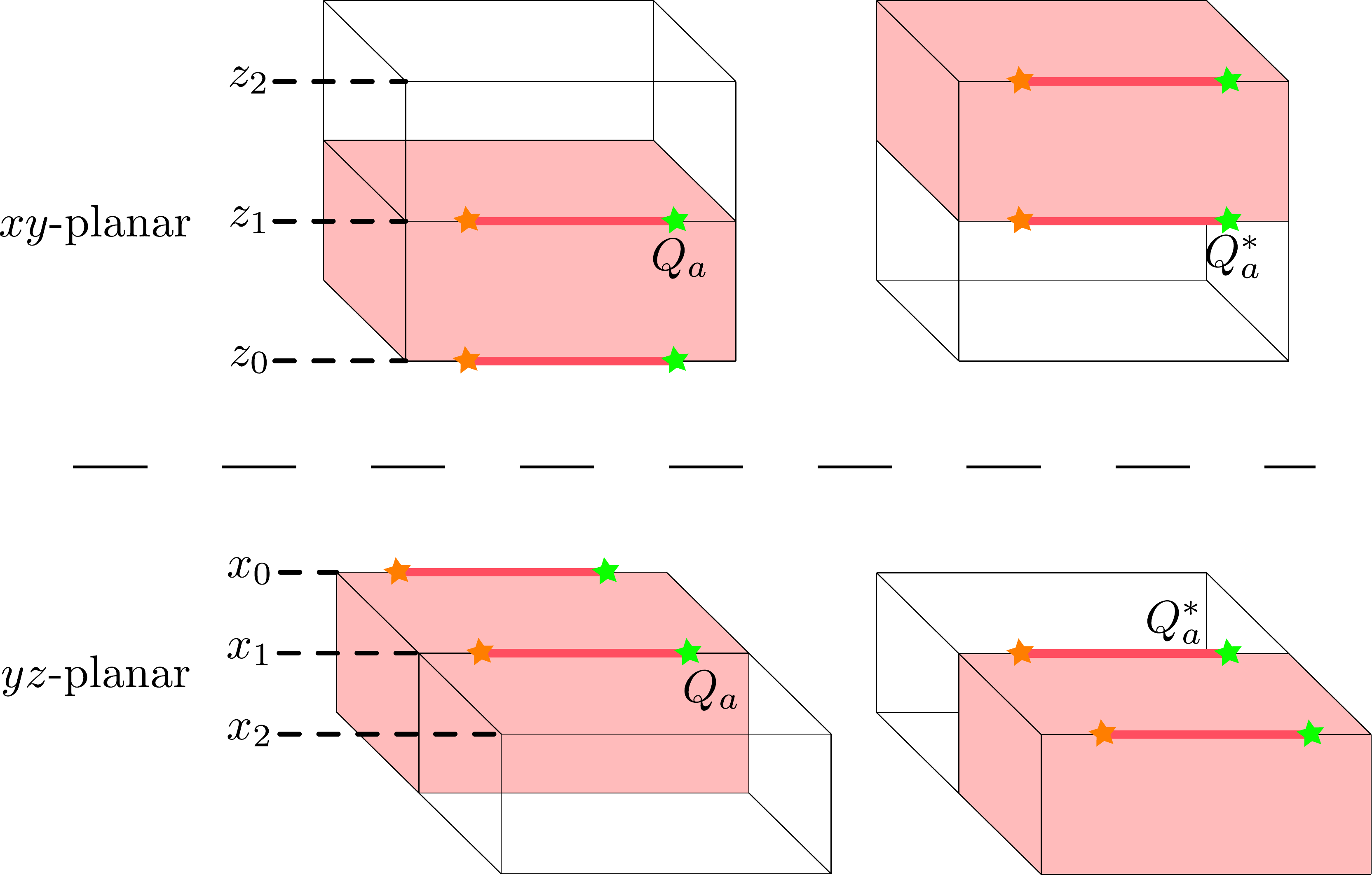}
\caption{
A truncated symmetry $S_M(g)$ (shown in pink) acts non-trivially along its hinges.  
Truncating this to a segment as in the Else-Nayak procedure allows us to identify the diagonal operator $Q_a$.
(top) The operator $Q_a$ is used to compute invariants $F_1$ and $F_2$ for $xy$-planar symmetries.
(bottom)
The same operator $Q_a$ is also used to compute invariants $F_1^{(yz)}$ and $F_2^{(yz)}$ using $yz$ planar symmetries.
}\label{fig:proofs}
\end{figure}

Now, let us begin talking about the SSPT.
Consider applying a symmetry $S_M(g)$ to a cubic region $M$, which acts non-trivially as $U_{h_M}(g)$ along the hinges.
Then, consider a symmetric truncation of $U_{h_M}(g)\rightarrow U_C(g)$ which leads to $V_a(g_1,g_2)$ in the Else-Nayak procedure, and consider $Q_a(x,z;g)$ on an upper hinge (see Fig.~\ref{fig:proofs}), 
 where we are now explicitly labeling the $x$ and $z$ coordinate of the hinge.
Notice that if we had instead chosen to consider $V_a^\prime(g_1,g_2)$ defined from the bottom hinge, we would end up with the conjugate $Q_a^*(x,z;g)$ instead (as shown in Fig.~\ref{fig:proofs}), which follows from the fact that the bottom hinge of $S_M(g)$ is related by a symmetry action to the top hinge of $S_M(g^{-1})$.
Knowing $Q_a(x,z;g)$ is sufficient to calculate the invariants $F_1$ and $F_2$.

Consider calculating $F_1$, using $H^3[G^L,U(1)]$ obtained from $xy$ planar symmetries.
Let us choose $g$ to be the generator of $G=\mathbb{Z}_{2N}$.
Then, the invariant $F_1$ corresponds to 
\begin{equation}
e^{\pi i F_1} = \Omega_{II}(g_<,g_\geq)^{N}
\end{equation}
where 
\begin{eqnarray}
g_< = \prod_{z=z_0}^{z_1-1} g_z\\
g_\geq = \prod_{z=z_1}^{z_2} g_z
\end{eqnarray}
for some  arbitrary $z_1$, with $z_0\ll z_1 \ll z_2$.
Let us take the region $M$ to be some region $x<x_1$, such that the relevant edge is at $x$-coordinate $x_1$.
Then, applying Eq~\ref{eq:omegacalc}
\begin{align}
\begin{split}
\Omega(g_<) &= \frac{\prescript{S_M(g_<)}{} Q_a^*(x_1,z_0; g)\prescript{S_M(g_<)}{} Q_a(x_1,z_1;g)}{Q_a^*(x_1,z_0;g) Q_a(x_1,z_1;g)}\\
\Omega(g_\geq) &= \frac{\prescript{S_M(g_\geq)}{} Q_a^*(x_1,z_1; g)\prescript{S_M(g_\geq)}{} Q_a(x_1,z_2;g)}{Q_a^*(x_1,z_1;g) Q_a(x_1,z_2;g)}\\
\Omega(g_< g_\geq) &= \frac{\prescript{S_M(g_< g_\geq)}{} Q_a^*(x_1,z_0; g)\prescript{S_M(g_< g_\geq)}{} Q_a(x_1,z_2;g)}{Q_a^*(x_1,z_0;g) Q_a(x_1,z_2;g)}
\end{split}
\label{eq:f1eqs}
\end{align}

\newcommand{\sqnw}[1]{ \begin{tikzpicture}[#1] \fill[color=gray] (0ex,0.75ex) rectangle (0.75ex,1.5ex); \draw (0,0) -- (1.5ex,0ex); \draw (1.5ex,0ex) -- (1.5ex,1.5ex); \draw (1.5ex,1.5ex) -- (0ex,1.5ex); \draw (0ex,1.5ex) -- (0ex,0ex); \draw (0.75ex,0ex) -- (0.75ex,1.5ex); \draw (0ex,0.75ex) -- (1.5ex,0.75ex); \end{tikzpicture} }
\newcommand{\sqne}[1]{ \begin{tikzpicture}[#1] \fill[color=gray] (0.75ex,0.75ex) rectangle (1.5ex,1.5ex); \draw (0,0) -- (1.5ex,0ex); \draw (1.5ex,0ex) -- (1.5ex,1.5ex); \draw (1.5ex,1.5ex) -- (0ex,1.5ex); \draw (0ex,1.5ex) -- (0ex,0ex); \draw (0.75ex,0ex) -- (0.75ex,1.5ex); \draw (0ex,0.75ex) -- (1.5ex,0.75ex); \end{tikzpicture} }
\newcommand{\sqsw}[1]{ \begin{tikzpicture}[#1] \fill[color=gray] (0ex,0ex) rectangle (0.75ex,0.75ex); \draw (0,0) -- (1.5ex,0ex); \draw (1.5ex,0ex) -- (1.5ex,1.5ex); \draw (1.5ex,1.5ex) -- (0ex,1.5ex); \draw (0ex,1.5ex) -- (0ex,0ex); \draw (0.75ex,0ex) -- (0.75ex,1.5ex); \draw (0ex,0.75ex) -- (1.5ex,0.75ex); \end{tikzpicture} }
\newcommand{\sqse}[1]{ \begin{tikzpicture}[#1] \fill[color=gray] (0.75ex,0ex) rectangle (1.5ex,0.75ex); \draw (0,0) -- (1.5ex,0ex); \draw (1.5ex,0ex) -- (1.5ex,1.5ex); \draw (1.5ex,1.5ex) -- (0ex,1.5ex); \draw (0ex,1.5ex) -- (0ex,0ex); \draw (0.75ex,0ex) -- (0.75ex,1.5ex); \draw (0ex,0.75ex) -- (1.5ex,0.75ex); \end{tikzpicture} }
\newcommand{\sqn}[1]{ \begin{tikzpicture}[#1] \fill[color=gray] (0ex,0.75ex) rectangle (1.5ex,1.5ex); \draw (0,0) -- (1.5ex,0ex); \draw (1.5ex,0ex) -- (1.5ex,1.5ex); \draw (1.5ex,1.5ex) -- (0ex,1.5ex); \draw (0ex,1.5ex) -- (0ex,0ex); \draw (0.75ex,0ex) -- (0.75ex,1.5ex); \draw (0ex,0.75ex) -- (1.5ex,0.75ex); \end{tikzpicture} }
\newcommand{\sqe}[1]{ \begin{tikzpicture}[#1] \fill[color=gray] (0.75ex,0ex) rectangle (1.5ex,1.5ex); \draw (0,0) -- (1.5ex,0ex); \draw (1.5ex,0ex) -- (1.5ex,1.5ex); \draw (1.5ex,1.5ex) -- (0ex,1.5ex); \draw (0ex,1.5ex) -- (0ex,0ex); \draw (0.75ex,0ex) -- (0.75ex,1.5ex); \draw (0ex,0.75ex) -- (1.5ex,0.75ex); \end{tikzpicture} }
\newcommand{\sqs}[1]{ \begin{tikzpicture}[#1] \fill[color=gray] (0ex,0ex) rectangle (1.5ex,0.75ex); \draw (0,0) -- (1.5ex,0ex); \draw (1.5ex,0ex) -- (1.5ex,1.5ex); \draw (1.5ex,1.5ex) -- (0ex,1.5ex); \draw (0ex,1.5ex) -- (0ex,0ex); \draw (0.75ex,0ex) -- (0.75ex,1.5ex); \draw (0ex,0.75ex) -- (1.5ex,0.75ex); \end{tikzpicture} }
\newcommand{\sqw}[1]{ \begin{tikzpicture}[#1] \fill[color=gray] (0ex,0ex) rectangle (0.75ex,1.5ex); \draw (0,0) -- (1.5ex,0ex); \draw (1.5ex,0ex) -- (1.5ex,1.5ex); \draw (1.5ex,1.5ex) -- (0ex,1.5ex); \draw (0ex,1.5ex) -- (0ex,0ex); \draw (0.75ex,0ex) -- (0.75ex,1.5ex); \draw (0ex,0.75ex) -- (1.5ex,0.75ex); \end{tikzpicture} }
\newcommand{\sqtot}[1]{ \begin{tikzpicture}[#1] \fill[color=gray] (0ex,0ex) rectangle (1.5ex,1.5ex); \draw (0,0) -- (1.5ex,0ex); \draw (1.5ex,0ex) -- (1.5ex,1.5ex); \draw (1.5ex,1.5ex) -- (0ex,1.5ex); \draw (0ex,1.5ex) -- (0ex,0ex); \draw (0.75ex,0ex) -- (0.75ex,1.5ex); \draw (0ex,0.75ex) -- (1.5ex,0.75ex); \end{tikzpicture} }
\newcommand{\sqdiag}[1]{ \begin{tikzpicture}[#1] \fill[color=gray] (0ex,0.75ex) rectangle (0.75ex,1.5ex); 
\fill[color=gray] (0.75ex,0ex) rectangle (1.5ex,0.75ex);
\draw (0,0) -- (1.5ex,0ex); \draw (1.5ex,0ex) -- (1.5ex,1.5ex); \draw (1.5ex,1.5ex) -- (0ex,1.5ex); \draw (0ex,1.5ex) -- (0ex,0ex); \draw (0.75ex,0ex) -- (0.75ex,1.5ex); \draw (0ex,0.75ex) -- (1.5ex,0.75ex); \end{tikzpicture} }

\begin{figure}
\includegraphics[width=0.3\textwidth]{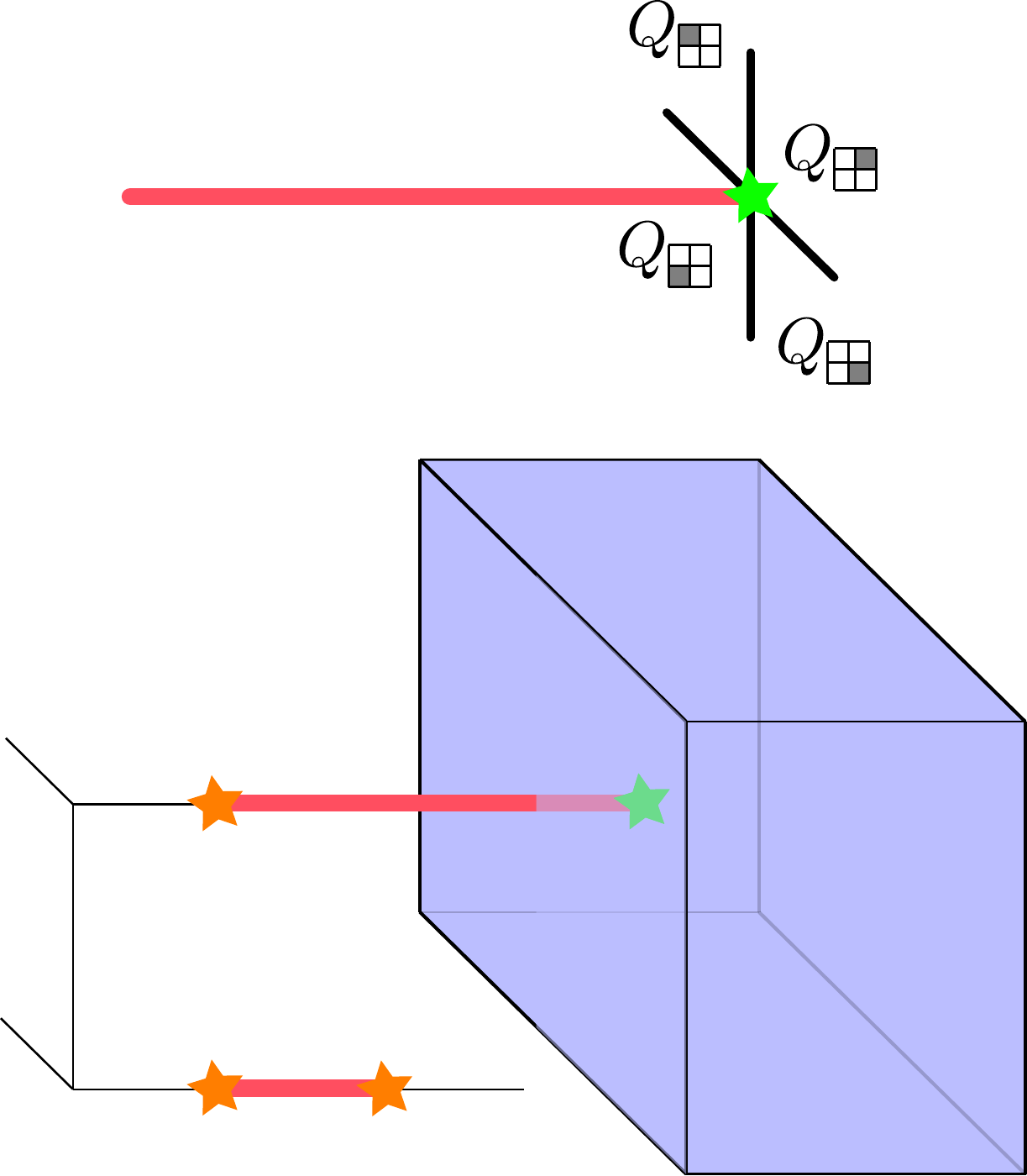}
\caption{
(top) The charge of the operator $Q_a$ is divided into contributions from four quadrants as shown.
(bottom) In a 3-foliated model, a product of $zx$-planar symmetries (blue) act as a global symmetry near $Q_a$ (the green star), while not acting on any of the other truncation points (orange stars).  
This implies that the total charge of $Q_a$ must be zero.  
}\label{fig:proofs2}
\end{figure}

For convenience, let us divide $Q_a(x_1,z_1;g)$ into four quadrants, as shown in Fig.~\ref{fig:proofs2}, and denote its charge in each quadrant as $Q_{\sqnw{}}$,$Q_{\sqne{}}$,$Q_{\sqse{}}$, and $Q_{\sqsw{}}$.
For example,
\begin{equation}
Q_{\sqsw{}} = \frac{\prescript{S_M(g_<)}{} Q_a(x_1,z_1;g)}{Q_a(x_1,z_1;g)}
\end{equation}
Using this, we can express using Eqs~\ref{eq:f1eqs}
\begin{equation}
\Omega_{II}(g_<,g_\geq) = Q_{\sqnw{}} / Q_{\sqsw{}} \label{eq:chargedist1}
\end{equation}

Alternatively, we could have used Eq.~\ref{eq:omega2calc1} and Eq.~\ref{eq:omega2calc2} to obtain
\begin{equation}
\Omega_{II}(g_<,g_\geq) = Q_{\sqn{}}  \label{eq:chargedist2}
\end{equation}
and 
\begin{equation}
\Omega_{II}(g_<,g_\geq) = 1/Q_{\sqs{}}  \label{eq:chargedist3}
\end{equation}
where $Q_{\sqn{}}\equiv Q_{\sqnw{}} Q_{\sqne{}}$, and similarly for others.

Eq.~\ref{eq:chargedist1},~\ref{eq:chargedist2}, and~\ref{eq:chargedist3} imply that the charge distribution in $Q$ must satisfy
\begin{equation}
1 = Q_{\sqnw{}}/Q_{\sqse{}} = Q_{\sqne{}}/Q_{\sqsw{}} 
\end{equation}
Thus, there are two degrees of freedom for the charge distribution in $Q$, which we may call $q_1$ and $q_2$, 
\begin{align}
\begin{split}
e^{2\pi i q_1 / (2N)} = Q_{\sqnw{}} = 1/Q_{\sqse{}}\\
e^{2\pi i q_2 / (2N)} = Q_{\sqne{}} = 1/Q_{\sqsw{}}\\
\end{split}
\end{align}
in which case $\Omega_{II}(g_<,g_{\geq}) = e^{2\pi i (q_1+q_2) / (2N)}$.
The invariant $F_1$ is then $F_1 = q_1 + q_2 \mod 2$.

Now, suppose we calculate the same quantity except using $yz$ planar symmetries instead.  
We may perform the calculation using the same hinge $Q_a(x_1,z_1;g)$, as shown in Fig.~\ref{fig:proofs}.
In this case, one finds that
\begin{equation}
\Omega_{II}^{(yz)}(g^{(yz)}_<,g^{(yz)}_{\geq}) = Q_{\sqe{}}  = e^{2\pi i (q_2-q_1)/(2N)}
\end{equation}
where we have explicitly labeled everything with $yz$ to avoid confusion ($g^{(yz)}_<$ is the product of $g^{(yz)}_x$ for $x<x_1$, for example).
In this case, one has $F_1^{(yz)} = q_2-q_1\mod 2$.
However, $q_2-q_1 = q_2 + q_1\mod 2$, and so $F_1^{(yz)} = F_1$ is independent of whether we had chosen the $xy$ or $yz$ plane.
In the 3-foliated case, we may use the same argument along a different hinge to show that $F_1^{(zx)}$ is also given by the same quantity.

Next, consider the quantity $F_2$.  Take $G=\mathbb{Z}_N\times\mathbb{Z}_N$, and choose $g$ and $h$ to be the two generators of $G$.
Then, we wish to compute
\begin{equation}
e^{2\pi i F_2/N} = \Omega_{II}(g_<, h_\geq) / \Omega_{II}(h_<,g_\geq)
\end{equation}
using the same set-up as before.
Let us define
\begin{equation}
Q_{\sqsw{}}^{h,g} = \frac{\prescript{S_M(h_<)}{} Q_a(x_1,z_1;g)}{Q_a(x_1,z_1;g)}
\end{equation}
to be the $h$ charge in the $\sqsw{}$ quadrant of $Q_a(x_1,z_1,g)$, and similarly for the other quadrants.
Then, using Eq.~\ref{eq:omega2calc1} and Eq.~\ref{eq:omega2calc2},
\begin{align}
\Omega_{II}(g_<,h_\geq) = Q^{h,g}_{\sqn{}}\\
\Omega_{II}(h_<,g_\geq) = 1/Q^{h,g}_{\sqs{}}
\end{align}
such that
\begin{equation}
e^{2\pi i F_2/N} = Q^{h,g}_{\sqtot{}}
\end{equation}
is simply the total $h$ charge of $Q_a(x_1,z_1,g)$.

Clearly, if we were to perform this calculation for the $yz$ plane using this same hinge ($x_1,z_1$), we would find exactly the same result, $e^{2\pi i F_2^{(yz)}/N}  = Q^{h,g}_{\sqtot{}}$.
(depending on choice of convention: we could have had $F_2^{(yz)}=-F_2$ instead).
Thus, $F_2$ is independent of whether we measure using the $xy$ or $yz$ planes.

Now, suppose our model is 3-foliated.
We have shown that if we consider every endpoint (not just $Q_a$), the total charge must be zero under any untruncated symmetry (Eq.~\ref{eq:zerocharge}).
However, in a 3-foliated model, we may choose a symmetry operator which acts like a global symmetry near $Q_a$, but does not act on the other endpoints at all (see Fig.~\ref{fig:proofs2}).
This means that $Q_a$ must have trivial total charge under any global symmetry.
Thus, 
$ e^{2\pi i F_2/N} = Q^{h,g}_{\sqtot{}} = 1 $
must be trivial.

On the gauged side this has a natural interpretation: for the gauged 3-foliated model, $N$ lineons at $(x_1,z_1)$ (which are mobile in the $y$ direction) cannot carry any charge under $S^{(zx)}(y;h)$, otherwise they could not have been mobile in the $y$ direction in the first place.


\subsection{Completeness and basis change}
Here, we first go through how to obtain $\matr{M}$ for a 3D SSPT after stacking by a 2D SPT, as described by the main text.
Then, we prove that the invariants $F_1$ and $F_2$ are a complete classification of all matrices $\matr{M}$ modulo this stacking. 
We prove this by showing that all possible $\matr{M}$ may be brought into a canonical form $\matr{M}^{\mathrm{canon}}$, determined solely by $F_1$ and $F_2$, via stacking 2D SPTs.

\subsubsection{Basis change}
First, let us go over the details of the basis change.
For this section, we will work with general $G=\prod_{\alpha=1}^{M} \mathbb{Z}_{N_\alpha}$.
In this case, $\matr{M}$ is an $M L \times M L$ integer matrix, where the off-diagonal elements
$M_{(\alpha,z),(\beta,z^\prime)}$ are defined modulo $\gcd(N_\alpha,N_\beta)\equiv N_{\alpha\beta}$, and the diagonal elements $M_{(\alpha,z),(\alpha,z)}$ are even integers modulo $2 N_\alpha$.

Recall that we wish to stack a 2D SPT with symmetry group $G_{2D}=G^K$, and that we do so by identifying each factor of $G$ in $G_{2D}$ with a plane $z_k$ in the SSPT.  
That is, let $k=1,\dots,K$ label the factors of $G$ in $G_{2D}$, which we associate with the plane $z_k$, and $\widetilde{M}^{2D}_{(\alpha,k),(\beta,k^\prime)}$ the $KM \times KM$ matrix characterizing the 2D SPT.
To ensure locality, all $\{z_k\}$ must reside within some finite $\mathcal{O}(1)$ interval.
Then, define $\matr{M}^{2D}$ to be the matrix with the same dimensions as $\matr{M}$, whose elements are obtained directly from $\widetilde{\matr{M}}^{2D}$,
\begin{equation}
M^{2D}_{(\alpha,z_k),(\beta,z_{k^\prime})} = \widetilde{M}^{2D}_{(\alpha,k),(\beta,k^\prime)}
\end{equation}
and all other elements with $z\notin \{z_k\}$ zero.
If we were stacking on to a 1-foliated model described by $\matr{M}$, we would simply modify $\matr{M}\rightarrow \matr{M}+\matr{M}^{2D}$.
However, when stacking to a 2- or 3-foliated model, we instead define the 2D SPT in terms of $d_\ve{r}$ degrees of freedom.
Thus, one instead has
$\matr{M}\rightarrow \matr{M} + \matr{W}^T\matr{M}^{2D}\matr{W}$,
where
\begin{equation}
W_{(\alpha,z),(\beta,z^\prime)} = \delta_{\alpha \beta} (\delta_{z+1,z^\prime}-\delta_{z,z^\prime})
\end{equation}
For example, suppose $G=\mathbb{Z}_N$ and we have a single type-I cocycle on plane $z_1$,
\begin{equation}
\matr{M}^{2D}=
\begin{blockarray}{cc}
z_1 & z_1+1\\
\begin{block}{[cc]}
  2 & 0\\
  0 & 0\\
\end{block}
\end{blockarray}
\end{equation}
where we show only the $\{z_1,z_1+1\}$ submatrix.
Then, within this submatrix,
\begin{align}
\begin{split}
\matr{W}^T\matr{M}^{2D}\matr{W} =&
\begin{bmatrix}
-1&0\\
1&-1
\end{bmatrix} 
\begin{bmatrix}
2&0\\
0&0
\end{bmatrix}
\begin{bmatrix}
-1&1\\
0&-1
\end{bmatrix}
\\=&
\begin{bmatrix}
2&-2\\
-2&2
\end{bmatrix} 
\end{split}
\end{align}
and all other elements outside of this submatrix are $0$.
This therefore results in two type-I cocycles valued $1$ (recall that the diagonal elements are $M_{ii}=2p_I^{i}$) on planes $z_1$ and $z_1+1$, and a type-II cocycle valued $-2$ between the two planes.
This is one of the examples shown in Fig.~1 of the main text.

\subsubsection{Completeness}
Next, let us show completeness of the invariants $F_1$ and $F_2$.
In a general group $G=\prod_{\alpha=1}^M \mathbb{Z}_{N_\alpha}$, we may define $F_1^{\alpha}$ for each even $N_\alpha$, and $F_2^{\alpha \beta}$ for each $N_{\alpha \beta}\neq 1$.
These are defined in reference to some plane $z_0$,
\begin{equation}
F_1^\alpha\equiv \sum_{z<z_0} \sum_{z^\prime \geq z_0} M_{(\alpha,z) (\alpha,z^\prime)}  \mod 2
\end{equation}
and
\begin{equation}
F_2^{\alpha\beta} \equiv\sum_{z<z_0}\sum_{z^\prime\geq z_0} \left(M_{(\alpha,z),(\beta,z^\prime)} - M_{(\beta,z),(\alpha,z^\prime)}\right)\mod N_{\alpha\beta}
\end{equation}
and are independent of the precise choice $z_0$.
Let us define equivalence classes of $\matr{M}$, where $\matr{M}_1$ and $\matr{M}_2$ belong to the same equivalence class if $\matr{M}_1 = \matr{M}_2 + \matr{W}^T\matr{M}^{2D}\matr{W}$ for some $\matr{M}^{2D}$.
Then, we claim that $F_1^\alpha$ and $F_2^{\alpha\beta}$ (for $\alpha<\beta$) are a complete set of invariants, and are sufficient to characterize all equivalence classes of $\matr{M}$.

Our strategy is as follows: assume we are given a general $\matr{M}$, with some longest range coupling $d_{max}$, defined as the maximum $|z-z^\prime|$ where $M_{(\alpha,z),(\beta,z^\prime)}$ is non-zero.
We then show that by stacking $\matr{M}^{2D}$ we can reduce $\matr{M}$ down to one in which $d_{max}=1$, such that there is only couplings between planes $z$ and $z\pm 1$.  
Then, we finally reduce $\matr{M}$ down to some canonical $\matr{M}^{\mathrm{canon}}$, which only depends on $F_1^\alpha$ and $F_2^{\alpha\beta}$.  
Thus, any two $\matr{M}$ with the same $F_1^\alpha$ and $F_2^{\alpha\beta}$ can be related to one another by stacking various $\matr{M}^{2D}$, and are therefore a complete set of invariants.

First, suppose we have some matrix $\matr{M}$ with some longest range coupling $d_{max}>1$ (which is always $\mathcal{O}(1)$ due to locality).
This means there is some element $M_{(\alpha_1,z_1),(\alpha_2,z_2)} \neq 0$ where $|z_2-z_1|=d_{max}$.
By symmetry of $\matr{M}$, we may consider $z_2>z_1$ without loss of generality.

Suppose $\alpha_1=\alpha_2$, then take
\begin{equation}
\matr{M}^{2D}_{\alpha_1,\alpha_2}=
\begin{blockarray}{cccc}
z_1 & z_1+1 & z_2-1 & z_2 \\
\begin{block}{[cc|cc]}
  0 & 0 & 1 & 0   \\
  0 & 0 & 0 & 0   \\
  \BAhline 
  1 & 0 & 0 & 0   \\
  0 & 0 & 0 & 0   \\
\end{block}
\end{blockarray}
\end{equation}
where $\matr{M}^{2D}_{\alpha_1\alpha_2}$ is viewed as a matrix indexed by $z$, with fixed $\alpha_1,\alpha_2$, and we only show the relevant non-zero submatrix.
We then have
\begin{equation}
(\matr{W}^T\matr{M}^{2D}\matr{W})_{\alpha_1,\alpha_2}=
\begin{blockarray}{cccc}
\begin{block}{[cc|cc]}
  0 & 0 & 1 & -1   \\
  0 & 0 & -1 & 1   \\
  \BAhline 
  1 & -1 & 0 & 0   \\
  -1 & 1 & 0 & 0   \\
\end{block}
\end{blockarray}
\end{equation}
which has a single $-1$ as its $((\alpha_1,z_1),(\alpha_2,z_2))$th element (along with its symmetric partner), and all other elements are of $|z-z^\prime|<d_{max}$.
Thus, we may take
\begin{equation}
\matr{M}^\prime = \matr{M} + M_{(\alpha_1,z_1),(\alpha_2,z_2)} \matr{W}^T\matr{M}^{2D}\matr{W}
\end{equation}
which now has $M^{\prime}_{(\alpha_1,z_1),(\alpha_2,z_2)} = 0$.
Note that although in writing the submatrix we have assumed $d_{max}=z_2-z_1>2$, this also works for $d_{max}=2$.

If $\alpha_1\neq\alpha_2$, then we may instead use
\begin{equation}
\matr{M}^{2D}_{\alpha_1,\alpha_2}=
\begin{blockarray}{cccc}
z_1 & z_1+1 & z_2-1 & z_2 \\
\begin{block}{[cc|cc]}
  0 & 0 & 1 & 0   \\
  0 & 0 & 0 & 0   \\
  \BAhline 
  0 & 0 & 0 & 0   \\
  0 & 0 & 0 & 0   \\
\end{block}
\end{blockarray}
\end{equation}
such that
\begin{equation}
(\matr{W}^T\matr{M}^{2D}\matr{W})_{\alpha_1,\alpha_2}=
\begin{blockarray}{cccc}
\begin{block}{[cc|cc]}
  0 & 0 & 1 & -1   \\
  0 & 0 & -1 & 1   \\
  \BAhline 
  0 & 0 & 0 & 0   \\
  0 & 0 & 0 & 0   \\
\end{block}
\end{blockarray}
\end{equation}
again only has a $-1$ as its $((\alpha_1,z_1),(\alpha_2,z_2))$th element, and all other elements have range smaller than $d_{max}$.

We may repeat this on all non-zero elements of $\matr{M}$ with distance $d_{max}$, after which we end up with some matrix with $d_{max}^\prime < d_{max}$.  
We can repeat this process until we have $d_{max}=1$,
meaning $\matr{M}_{\alpha_1 \alpha_2}$ is a tridiagonal matrix.

Let us now define a canonical form $\matr{M}^{\mathrm{canon}}$, for a given set of $F_1^\alpha$ and $F_2^{\alpha\beta}$, by
\begin{equation}
\matr{M}^{\mathrm{canon}}_{\alpha \alpha} = 
\begin{blockarray}{cccccc}
\begin{block}{[cccccc]}
        \ddots&\ddots&&&&\\
        & -2 F_1^\alpha & F_1^\alpha & 0 & 0&   \\
        \ddots&F_1^\alpha & -2F_1^\alpha & F_1^\alpha & 0 &   \\
        &0 & F_1^\alpha & -2F_1^\alpha & F_1^\alpha &  \\
        &0 & 0 & F_1^\alpha & -2F_1^\alpha &\ddots \\
        &&&&\ddots&\ddots  \\
\end{block}
\end{blockarray}
\end{equation}
and 
\begin{equation}
\matr{M}^{\mathrm{canon}}_{\alpha \beta} = 
\begin{blockarray}{cccccc}
\begin{block}{[cccccc]}
        \ddots&\ddots&&&&\\
        & -F_2^{\alpha\beta} & F_2^{\alpha\beta} & 0 & 0&   \\
        &0 & -F_2^{\alpha\beta} & F_2^{\alpha\beta} & 0 &   \\
        &0 & 0 & -F_2^{\alpha\beta} & F_2^{\alpha\beta} &  \\
        &0 & 0 & 0 & -F_2^{\alpha\beta} &\ddots \\
        &&&&&\ddots  \\
\end{block}
\end{blockarray}
\end{equation}
for $\alpha<\beta$.  For $\beta<\alpha$, we simply have $\matr{M}^{\mathrm{canon}}_{\alpha\beta} = (\matr{M}^{\mathrm{canon}}_{\beta\alpha})^T$. 
We have also simply set $F_1^\alpha=0$ for any odd $N_{\alpha}$, and $F_2^{\alpha\beta}=0$ for any $N_{\alpha\beta}=1$.
The strong examples shown in Fig.~1 of the main text are both already in canonical form.
We will now show that our tridiagonal $\matr{M}$ can always be brought into its canonical form.

First, for each $\alpha$, examine the symmetric matrix $\matr{M}_{\alpha \alpha}$.
Consider each $2\times2$ block coupling $z_1$ and $z_1+1$.
We may stack with
\begin{equation}
\matr{M}^{2D}_{\alpha,\alpha}=
\begin{blockarray}{cc}
z_1 &z_1+1\\
\begin{block}{[cc]}
  2 & 0 \\
  0 & 0 \\
\end{block}
\end{blockarray}
\end{equation}
which realizes
\begin{equation}
(\matr{W}^T \matr{M}^{2D} \matr{W})_{\alpha,\alpha}=
\begin{blockarray}{cc}
\begin{block}{[cc]}
  2 & -2 \\
  -2 & 2 \\
\end{block}
\end{blockarray}
\end{equation}
which we can add to $\matr{M}$ to modify the offdiagonal element to be $0$ or $1$ depending on its parity (if $N_\alpha$ even) or $0$ (if $N_\alpha$ odd).
We may do this for all the offdiagonal elements, bringing them all to $F_1^\alpha$.
The diagonal elements are automatically constrained by the local constraint (Eq~\ref{eq:applocalconstraint}) to be $-2 F_1^\alpha$.

Next, we may do a similar thing to $\matr{M}_{\alpha,\beta}$ for each $\alpha<\beta$.
In this case, we stack
\begin{align}
\matr{M}^{2D}_{\alpha,\beta}&=
\begin{blockarray}{cc}
z_1 &z_1+1\\
\begin{block}{[cc]}
  1 & 0 \\
  0 & 0 \\
\end{block}
\end{blockarray}\\
(\matr{W}^T \matr{M}^{2D} \matr{W})_{\alpha,\beta}&=
\begin{blockarray}{cc}
\begin{block}{[cc]}
  1 & -1 \\
  -1 & 1 \\
\end{block}
\end{blockarray}
\end{align}
which we can use to eliminate all the lower-diagonal elements $M_{(\alpha,z_1+1),(\beta,z_1)}$.
Then, the upper-diagonal elements are $M_{(\alpha,z_1),(\beta,z_1+1)}=F_2^{\alpha\beta}$ and the diagonal elements are all automatically fixed by the local constraint to be $-F_2^{\alpha\beta}$.
We have therefore brought an arbitrary initial matrix $\matr{M}$, via moves of the form $\matr{W}^T\matr{M}^{2D}\matr{W}$ (stacking 2D SPTs), to a canonical form which only depends on $F_1^\alpha$ and $F_2^{\alpha\beta}$.
From this, we conclude that $F_1^\alpha$ and $F_2^{\alpha\beta}$ are a complete set of invariants for  $\matr{M}$.

\section{Strong models}\label{app:strongmodel}

In this section, we introduce two exactly solvable models of strong planar SSPT phases.
The first is the 3-foliated Type 1 strong phase with $G=\mathbb{Z}_2$, which we write down in the form of a Hamiltonian.
The fracton dual is a novel fracton model which we explicitly write down.
The second is the 2-foliated Type 1 and Type 2 strong phase with $G=\mathbb{Z}_N\times\mathbb{Z}_N$, for which we write down the ground state wavefunction $\ket{\psi}$. 
We may consider the 2-foliated model as part of a model with two sets of 2-foliated symmetries, in which case the fracton dual is again a novel model with unusual braiding statistics between fractons.
Alternatively, we may examine the fracton dual of a single 2-foliated model by itself, which results in a 2-foliated fracton phase, with non-trivial braiding statistics between gauge fluxes. 
To obtain models for strong phases for more general groups $G$, one may simply identify $\mathbb{Z}_2$ or $\mathbb{Z}_N\times\mathbb{Z}_N$ subgroups of $G$, and define the model in terms of those degrees of freedom.

\subsection{3-foliated Type 1 strong model}
The $G=\mathbb{Z}_2$ strong 3-foliated model is defined on the square lattice with qubit degrees of freedom on each site.
Define the Pauli matrices $Z$ and $X$,
\begin{align}
Z = \begin{bmatrix} 1&0\\0&-1\end{bmatrix}\;\;\;
X = \begin{bmatrix} 0&1\\1&0\end{bmatrix}
\end{align}
as well as the $S=\sqrt{Z}$ matrix and the controlled-Z ($CZ$) matrix
\begin{align}
S &= i^{(1-Z)/2} = \begin{bmatrix} 1&0\\0&i\end{bmatrix}\\
CZ_{12} &= (-1)^{(1-Z_1)(1-Z_2)/4} = 
\begin{bmatrix}
1&0&0&0\\
0&1&0&0\\
0&0&1&0\\
0&0&0&-1
\end{bmatrix}
\end{align}

The Hamiltonian will be written as a sum of terms of the form 
\begin{equation}
H = -\sum_\ve{r} X_\ve{r} F_\ve{r}(\{Z_p\}) \equiv -\sum_\ve{r} B_\ve{r}
\end{equation}
where $Z_p$ are products of $Z$ on the four corners of a plaquette $p$, and $F_\ve{r}(\{Z_p\})$ is some function of these variables near the site $\ve{r}$.
The planar symmetries will act as products of $X$s along $xy$, $yz$, or $zx$ planes.
As $F_\ve{r}(\{Z_p\})$ only depends on the combinations $Z_p$ which commutes with all planar symmetries, this Hamiltonian is explicitly symmetry respecting.  

\begin{figure}
\includegraphics[width=0.5\textwidth]{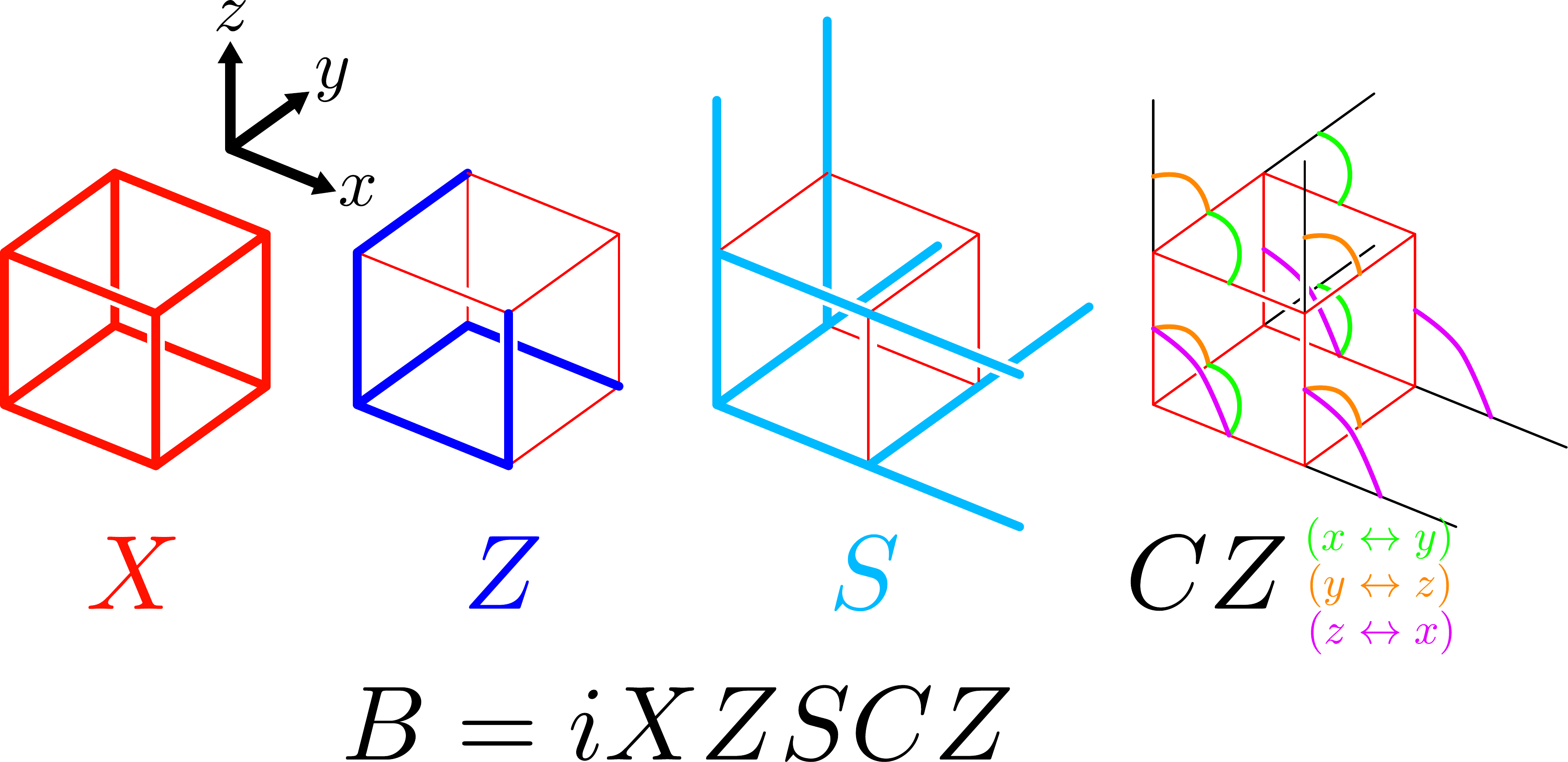}
\caption{
The operator $B_c$ for each cube in the gauged Type 1 strong model.
The action of the $Z$, $S$, and $CZ$ operators precede the action of the $X$.
The $CZ$ operators are always between two bonds oriented in different directions, and are denoted by a line connecting the two bonds.
For ease of viewing, $CZ$ operators between bonds of various pairs of orientations are shown in a different color.
The model is symmetric under three-fold rotation about the $(111)$ axis.
}\label{fig:z2strong}
\end{figure}

\begin{figure*}
\includegraphics[width=0.9\textwidth]{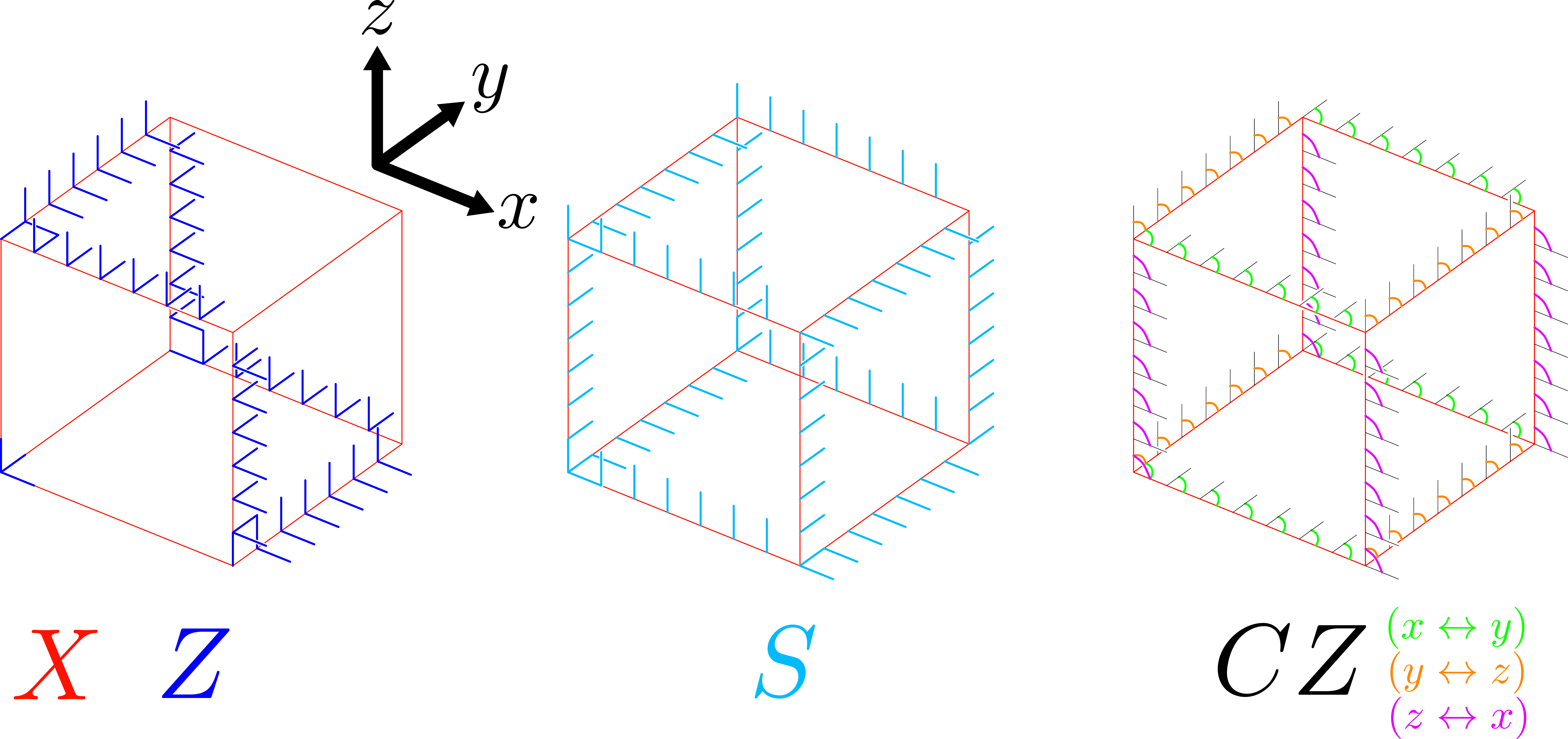}
\caption{
Taking a product of the cubic terms $B_c$ (Fig.~\ref{fig:z2strong}) results in a wireframe operator with support along the hinges of the cube.
This wireframe operator is shown here for a $7\times 7\times 7$ cube, where the action of the $Z$, $S$, and $CZ$ operators precede the $X$ (which acts along the red cube).
}\label{fig:wire}
\end{figure*}

The function $F_\ve{r}(\{Z_p\})$ consists of 6 $Z_p$, 12 $S_p$, and 12 $CZ_{p_1 p_2}$ operators on various plaquettes, and an overall factor of $i$.
Fig.~\ref{fig:z2strong} shows the model on the dual lattice, where plaquettes are represented by bonds, and the site $\ve{r}$ is mapped on to the red cube.
Careful calculation will show that $[B_\ve{r},B_{\ve{r}^\prime}]=0$  and $B_\ve{r}^2 = 1$.
This Hamiltonian is is therefore simply a commuting projector Hamiltonian, and as every term is independent (only $B_\ve{r}$ can act as $X_\ve{r}$) and there are the same number of terms as sites, $H$ has a unique gapped group state $\ket{\psi}$ and describes a valid SSPT.
We found it simplest to write a small computer script to confirm these commutation relations (and to compute the wireframe operator later), rather than doing so by hand.

The wireframe operator (Fig.~\ref{fig:wire}) obtained as a product of $B_\ve{r}$ over a large cube, when ungauged, gives the action of the symmetry on the hinges of the cube.
One may confirm using the Else-Nayak procedure that this model has $\matr{M}$ matrix
\begin{align}
\begin{split}
M_{z,z} &= 2\\
M_{z,z+1}&= M_{z,z-1} = 1
\end{split}
\end{align}
and all other elements zero.
This therefore realizes the Type 1 strong phase shown in Fig.~1 for $G=\mathbb{Z}_2$.
Computing the charges $Q$ defined in Section~\ref{app:proofs}, one finds $Q_{\sqnw{}} = Q_{\sqse{}} = -1$ and $Q_{\sqsw{}} = Q_{\sqne{}} = 1$.  

The fracton dual of this model is defined on the square lattice with qubit degrees of freedom on the bonds.
The Hamiltonian is given by
\begin{equation}
H =  - \sum_v (A_v^{xy} + A_v^{yz} + A_v^{zx}) - \sum_c B_{c}
\end{equation}
where $c$ represents cubes, $B_c$ is the operator shown in Fig.~\ref{fig:z2strong},
$v$ represents vertices, and $A_v^{\mu\nu}$ is the product of $Z$s along the four bonds touching $v$ in the $\mu\nu$ plane (the usual cross term from the X-cube model).
$B_c$ consists of $X$s along the cube (the cube term from the X-cube model) but with an additional phase factor depending on the $Z$ state around it in the form of $S$, $Z$, and $CZ$ operators.
Note that while the ungauged operator $B_\ve{r}$ squares to 1, the gauged operator $B_c$ does not square to 1, it instead squares to a product of $A_v$ operators.

This model has the same fracton charge excitations as the usual X-cube model.  
However, the lineon excitations are modified.
To find out what they are, consider the product of $B_c$ over a large cube, $\prod_{c} B_c$, shown in Fig.~\ref{fig:wire}.
This results in an operator with support only along the hinges of the cube.
This operator, when truncated, is the operator which creates lineon excitations at its ends.  

From this, the crossing (braiding) statistic of two lineon can be readily extracted.
Reading off of Fig.~\ref{fig:wire}, 
a pair of $x$-moving lineons on line $(y_1,z_1)$ is constructed by the operator
\begin{equation}
L_x \equiv \prod_{x=x_0}^{x_2} X_{x,y_1,z_1}^{(x)} S_{x,y_1,z_1}^{(z)} CZ_{x,y_1,z_1}^{(x\leftrightarrow y)}
\end{equation}
where $X_{x,y,z}^{(x)}$ is an $X$ on the bond originating from the vertex at $(x,y,z)$ going in the positive $x$ direction, and similarly for $S_{x,y,z}^{(z)}$, and $CZ_{x,y,z}^{(x\leftrightarrow y)}$ is a $CZ$ between $Z_{x,y,z}^{(x)}$ and $Z_{x,y,z}^{(y)}$.
$L_x$ creates two lineons at $x_0$ and $x_2$.
Meanwhile, a pair of $y$-moving lineons is constructed by
\begin{equation}
L_y \equiv \prod_{y=y_0}^{y_2} X_{x_1,y,z_1}^{(y)} S_{x_1,y,z_1}^{(x)} CZ_{x_1,y,z_1}^{(y\leftrightarrow z)}
\end{equation}
which creates two lineons at $y_0$ and $y_2$.
Note that depending on which hinge of the wireframe we obtain $L_x$ and $L_y$ from, there may be additional $Z$ operators, which correspond to a choice of lineon or antilineon (and will affect the braiding phase by a $\pm 1$).  
It can be readily verified that  when these two operators cross (i.e. $y_0<y_1<y_2$ and $x_0<x_1<x_2$), they only commute up to a factor of $i$,
\begin{equation}
L_y L_x = i L_y L_x
\end{equation}
using the relations $X S X = i Z S$ and $X_1 CZ_{12} X_1 = Z_2 CZ_{12}$.
Thus, the braiding phase of any two lineons in this model is $\pm i$.


\subsection{2-foliated strong model}
In this section, we describe a 2-foliated model which realizes both Type 1 and/or Type 2 strong phases.  
\subsubsection{A group cohomology model on the square lattice}
First, let us explicitly construct a group cohomology model on the square lattice, for $G=\mathbb{Z}_N^{M}$.
Recall that the ground state of such models are an equal amplitude sum of all configurations
\begin{equation}
\ket{\psi} = \sum_{\{g_\ve{r}\}} f(\{g_\ve{r}\})\ket{\{g_\ve{r}\}}
\end{equation}
where $f(\{g_\ve{r}\})$ is a pure phase up to an overall normalization, which we ignore.

From Eq~\ref{eq:groupcohomodel}, $f(\{g_\ve{r}\})$ is a product of terms $f_\ve{r}$ coming from each square plaquette at $\ve{r}$, given by
\begin{equation}
f_\ve{r}(\{g_\ve{r}\}) =
\frac{\nu(g_\ve{r}, g_{\ve{r}+\ve{x}}, g_{\ve{r}+\ve{x}+\ve{y}}, 1)}{\nu(g_\ve{r}, g_{\ve{r}+\ve{y}}, g_{\ve{r}+\ve{x}+\ve{y}}, 1)}
\end{equation}

Alternatively, we may choose to defined the same wavefunction using instead
\begin{equation}
f_\ve{r}^\prime(\{g_\ve{r}\}) =
\frac{\nu(1,g_\ve{r}, g_{\ve{r}+\ve{y}}, g_{\ve{r}+\ve{x}+\ve{y}})}{\nu(1,g_\ve{r}, g_{\ve{r}+\ve{x}}, g_{\ve{r}+\ve{x}+\ve{y}})}
\end{equation}
which one can verify using the cocycle condition (Eq~\ref{eq:nucocycle}) differs from $f_\ve{r}$ only by terms along the edges of the plaquette which are cancelled out by the same terms from neighboring plaquettes.
Plugging the explicit form for the cocycles, we get
\begin{widetext}
\begin{align}
\begin{split}
f_\ve{r}^{(2D)}(\{g_\ve{r}\}) = \exp\left\{
\sum_{i\leq j} \frac{2\pi i p^{ij}}{N^2}
g_\ve{r}^i 
\left(
[g^j_{\ve{r}+\ve{y}} - g^j_\ve{r}] + [g^j_{\ve{r}+\ve{x}+\ve{y}} - g^j_{\ve{r}+\ve{y}}] - [g^j_{\ve{r}+\ve{x}}-g^j_\ve{r}] - [g^j_{\ve{r}+\ve{x}+\ve{y}} - g^j_{\ve{r}+\ve{x}}]
\right)
\right\}
\end{split}
\end{align}\label{eq:f1}
\end{widetext}
which we have called $f_\ve{r}^{(2D)}$.

\subsubsection{The strong SSPT}
Let us take $G=\mathbb{Z}_N\times\mathbb{Z}_N$, with subsystem symmetries along $xy$ and $yz$ planes.
The ground state of our strong SSPT is again described by a function $f(\{g_\ve{r}\})$, which can be written as a product of $f_\ve{r}(\{g_\ve{r}\})$, which are now associated with the cube at $\ve{r}$.
The function $f_\ve{r}$ is given by
\begin{widetext}
\begin{equation}
\begin{alignedat}{2}
f_\ve{r}^{(SSPT)}(\{g_\ve{r}\}) = \exp
\left\{
\sum_{\alpha \leq \beta} \frac{2\pi i q^{\alpha \beta}}{N^2}
\left(
\vphantom{\sum_{\alpha\leq \beta} \frac{2\pi i p^{\alpha \beta}}{N^2}} 
\right.
\right.
&
(g^\alpha_{\ve{r}+\ve{z}} - g^\alpha_{\ve{r}})
(&&
[g^\beta_{\ve{r}+\ve{y}+\ve{z}} - g^\beta_{\ve{r}+\ve{z}}] + [g^\beta_{\ve{r}+\ve{x}+\ve{y}+\ve{z}} - g^\beta_{\ve{r}+\ve{y}+\ve{z}}]
\\
&&&- [g^\beta_{\ve{r}+\ve{x}+\ve{z}}-g^\beta_{\ve{r}+\ve{z}}] - [g^\beta_{\ve{r}+\ve{x}+\ve{y}+\ve{z}} - g^\beta_{\ve{r}+\ve{x}+\ve{z}}]
)\\
-&(g^\alpha_{\ve{r}+\ve{y}}-g^\alpha_\ve{r})
(&& [g^\beta_{\ve{r}+\ve{x}+\ve{y}+\ve{z}}-g^\beta_{\ve{r}+\ve{y}+\ve{z}}] - [g^\beta_{\ve{r}+\ve{x}+\ve{y}}-g^\beta_{\ve{r}+\ve{y}}]\\
&&& + [g^\beta_{\ve{r}+\ve{y}+\ve{z}}-g^\beta_{\ve{r}+\ve{x}+\ve{y}+\ve{z}} + g^\beta_{\ve{r}+\ve{x}+\ve{y}} - g^\beta_{\ve{r}+\ve{y}}] )
\left.
\left.
\vphantom{\sum_{\alpha\leq \beta} \frac{2\pi i p^{\alpha \beta}}{N^2}} 
\right)\right\}
\end{alignedat}\label{eq:strongf}
\end{equation}
\end{widetext}
for $q^{\alpha\beta}$ integers mod $N$.
Here, $\alpha,\beta\in\{1,2\}$ for each factor of $\mathbb{Z}_N$ in $G$.  

We claim that $f_\ve{r}^{(SSPT)}$ describes an SSPT phase which is Type 1 strong if $q^{11}$ or $q^{22}$ are odd (and $N$ is even), and Type 2 strong if $q^{12}\neq 0$.

First, let us examine the state as a quasi-2D SPT along the $xy$ plane, with a $G^L$ symmetry group.
Let us label each generator of $G^L$ by $(\alpha,z)$, for $\alpha\in\{1,2\}$ and $z\in[1,L]$.
The second term in the exponent (the term multiplying $(g^i_{\ve{r}+\ve{y}}-g_\ve{r}^i)$) is completely invariant under an $xy$ planar symmetry.  
This second term therefore cannot affect the $xy$ cocycle class, as it can be removed by a symmetric local unitary transformation respecting all $xy$ planar symmetries (but will break the $yz$ planar symmetries).  
Thus, the $xy$ cocycle class is determined simply by the first term.
However, this term is exactly of the form $f^{(2D)}_\ve{r}(\{g_\ve{r}\})$ for $G^L$, with the mapping
\begin{align}
\begin{split}
p^{(\alpha,z),(\beta,z)} &= q^{\alpha\beta}\\
p^{(\alpha,z),(\beta,z+1)} &= -q^{\alpha\beta}
\end{split}
\end{align}
and other elements zero.
In terms of the $\matr{M}$ matrix,
\begin{align}
\begin{split}
M_{(\alpha,z),(\beta,z)} &= (1+\delta_{\alpha \beta})q^{\alpha \beta}\\
M_{(\alpha,z),(\beta,z+1)} &= -q^{\alpha \beta}\\
\end{split}
\end{align}
and all other elements (except those related by symmetry) are zero.

The $F_1$ invariants are therefore simply $q^{11}$ and $q^{22}$ modulo 2, and the $F_2$ invariant is $-q^{12}$.
By the proof from our previous section, the invariants will also the same for the $yz$ symmetries.

But before we can conclude that we have constructed a strong phase, we must show that this state is symmetric under $yz$ symmetries.
The purpose of the second term in $f^{(SSPT)}_\ve{r}$ is to ensure that this is the case. 
Let us examine how $f_\ve{r}(\{g_\ve{r}\})$ transforms under a $yz$ planar symmetry which sends
$\{g_\ve{r}\}\rightarrow\{g^{(yz)}g_\ve{r}\}$, or, on the relevant degrees of freedom,
\begin{flalign}
\begin{split}
 &(g_{\ve{r}},g_{\ve{r}+\ve{y}}, g_{\ve{r}+\ve{z}},g_{\ve{r}+\ve{y}+\ve{z}})\rightarrow
 (gg_{\ve{r}},gg_{\ve{r}+\ve{y}}, gg_{\ve{r}+\ve{z}},gg_{\ve{r}+\ve{y}+\ve{z}})\\
 &(g_{\ve{r}+\ve{x}},g_{\ve{r}+\ve{x}+\ve{y}}, g_{\ve{r}+\ve{x}+\ve{z}},g_{\ve{r}+\ve{x}+\ve{y}+\ve{z}})\;\;\;\;\mathrm{unchanged}
\end{split}\label{eq:yzsymac}
\end{flalign}
A calculation shows that
\begin{widetext}
\begin{equation}
\begin{alignedat}{2}
\frac{f_\ve{r}(\{g^{(yz)} g_\ve{r}\})}{f_\ve{r}(\{g_\ve{r}\})} 
= \exp
\left\{
\sum_{\alpha \leq \beta} \frac{2\pi i q^{\alpha \beta}}{N^2}
\left(
\vphantom{\sum_{\alpha\leq \beta} \frac{2\pi i p^{\alpha \beta}}{N^2}} 
\right.
\right.
&
(g^\alpha_{\ve{r}+\ve{z}} - g^\alpha_{\ve{r}})
(&&
[g^\beta_{\ve{r}+\ve{x}+\ve{y}+\ve{z}} - g^\beta_{\ve{r}+\ve{y}+\ve{z}} + g^\beta]
-[g^\beta_{\ve{r}+\ve{x}+\ve{y}+\ve{z}} - g^\beta_{\ve{r}+\ve{y}+\ve{z}}]
\\
&&&- [g^\beta_{\ve{r}+\ve{x}+\ve{z}}-g^\beta_{\ve{r}+\ve{z}}+g^\beta] 
+ [g^\beta_{\ve{r}+\ve{x}+\ve{z}}-g^\beta_{\ve{r}+\ve{z}}] 
)\\
-&(g^\alpha_{\ve{r}+\ve{y}}-g^\alpha_\ve{r})
(&& [g^\beta_{\ve{r}+\ve{x}+\ve{y}+\ve{z}}-g^\beta_{\ve{r}+\ve{y}+\ve{z}}+g^\beta] 
-[g^\beta_{\ve{r}+\ve{x}+\ve{y}+\ve{z}}-g^\beta_{\ve{r}+\ve{y}+\ve{z}}]\\
&&&-[g^\beta_{\ve{r}+\ve{x}+\ve{y}}-g^\beta_{\ve{r}+\ve{y}}+g^\beta]
+[g^\beta_{\ve{r}+\ve{x}+\ve{y}}-g^\beta_{\ve{r}+\ve{y}}])
\left.
\left.
\vphantom{\sum_{\alpha\leq \beta} \frac{2\pi i p^{\alpha \beta}}{N^2}} 
\right)\right\}
\end{alignedat}
\end{equation}
which simplifies to
\begin{equation}
\frac{f_\ve{r}(\{g^{(yz)} g_\ve{r}\})}{f_\ve{r}(\{g_\ve{r}\})} 
= \frac{P(g_{\ve{r}+\ve{z}}, g_{\ve{r}+\ve{x}+\ve{y}+\ve{z}}, g_{\ve{r}+\ve{y}+\ve{z}}, g)}{
 P(g_{\ve{r}}, g_{\ve{r}+\ve{x}+\ve{y}}, g_{\ve{r}+\ve{y}}, g)}
 \frac{P(g_{\ve{r}}, g_{\ve{r}+\ve{x}+\ve{z}}, g_{\ve{r}+\ve{z}}, g)}{
 P(g_{\ve{r}+\ve{y}}, g_{\ve{r}+\ve{x}+\ve{y}+\ve{z}}, g_{\ve{r}+\ve{y}+\ve{z}}, g)}
 \frac{P(g_{\ve{r}+\ve{y}}, g_{\ve{r}+\ve{x}+\ve{y}}, g_{\ve{r}+\ve{y}}, g)}{
 P(g_{\ve{r}+\ve{z}}, g_{\ve{r}+\ve{x}+\ve{z}}, g_{\ve{r}+\ve{z}}, g)}
\end{equation}
where 
\begin{equation}
P(g_1,g_2,g_3,g) = \exp\left\{\sum_{\alpha\leq\beta}\frac{2\pi i q^{\alpha\beta}}{N^2}\left(g_1^\alpha([g_2^\beta-g_3^\beta+g^\beta]-[g_2^\beta-g_3^\beta])\right)\right\}
\end{equation}
\end{widetext}
If one considers the contribution from neighboring cubes, one finds that the factors of $P(\dots)$ exactly cancel out between neighboring cubes.
Repeating this calculation for a $yz$-planar symmetry which transforms the other four sites in Eq~\ref{eq:yzsymac}, one finds the same result.
Thus, the wavefunction is indeed symmetric under $yz$ planar symmetries and describes a strong SSPT phase for a 2-foliated model.
If one wished, one could confirm that the matrix $\matr{M}^{(yz)}$ obtained from $yz$ planar is also strong with the same $F_1$ and $F_2$ invariants, by following the Else-Nayak procedure.
Obtaining a gapped local Hamiltonian corresponding to this ground state is straightforward, and is done in the same way as for the standard group cohomology models, Eq~\ref{eq:gcham}.

\section{Relation to $p$-string condensation}\label{app:coupledlayer}

In this section, we will discuss the gauge duality between weak 3-foliated SSPTs and twisted X-cube models constructed by stacking 1-foliated gauge theories and inducing a $p$-string condensation transition.\cite{coupledlayer,coupledlayer2}
This procedure is a straightforward generalization of the coupled layers construction of the X-cube model and its twisted variants. We will demonstrate the correspondence by showing that our zero-correlation length Hamiltonian models for weak SSPTs are dual to the effective Hamiltonians that emerge from strongly coupling stacked 1-foliated gauge theories.

First, let us briefly view the coupled layers construction of the X-cube model. The starting point is 3 intersecting stacks of 2D toric code layers, oriented along $xy$, $yz$, and $zx$ planes respectively. The toric code layers contain qubits on the edges of square lattices, subject to the Hamiltonian
\begin{equation}
    H_{TC}=-\sum_v A_v-\sum_p B_p
\end{equation}
where $A_v$ is the tensor product of Pauli $Z$ operators over the four edges adjacent to vertex $v$, and $B_p$ is the product of Pauli $X$ operators over the four edges around plaquette $p$. They are arranged such that the degrees of freedom coincide on the edges of a cubic lattice, such that each edge contains two qubits from different stacks. Then, the stacks are coupled together via the term $ZZ$ acting on the two qubits on a given edge. In the strong-coupling limit, the two qubit degrees of freedom become one effective qubit degree of freedom, and the X-cube Hamiltonian
\begin{equation}
    H_{XC}=-\sum_v (A_v^{xy}+A_v^{yz}+A_v^{zx})-\sum_c B_c
\end{equation}
emerges as the effective Hamiltonian
governing these degrees of freedom.\cite{coupledlayer,coupledlayer2}
Here $A_v^{\mu\nu}$ is the tensor product of Pauli $Z$ operators over the edges emanating from vertex $v$ in the $\mu\nu$ plane, and $B_c$ is the product of $X$ operators over the 12 edges of the elementary cube $c$.
The transition between decoupled stacks and emergent X-cube order is driven by a mechanism which has been named $p$-string condensation.~\cite{coupledlayer} The essential idea is that the $ZZ$ coupling acts on the ground state by creating a small loop composed of 4 charge excitations. In the strong coupling limit, loops of gauge charges of all sizes, called $p$-strings, proliferate throughout the system and form a condensate in the ground state.

This construction can be generalized by replacing each stack of 2D toric codes by an arbitrary 1-foliated gauge theory, viewed as a quasi-2D topological order that compactifies to a $G^L$ twisted gauge theory (with $G$ an arbitrary abelian group). An example of such a 1-foliated gauge theory was constructed in Ref.~\onlinecite{foliated5}.
Exactly solvable Hamiltonians for 1-foliated gauge theories can be constructed as generalized string-net models. For simplicity let us consider the case $G=\mathbb{Z}_2$, with planar symmetries oriented along the $xy$ plane; the generalization to arbitrary abelian gauge group is straightforward. The degrees of freedom are qubits attached to the $x$- and $y$-oriented links of a cubic lattice, and the Hamiltonian for an arbitrary 1-foliated gauge theory takes the general form
\begin{equation}
    H_\textrm{1-fol}=-\sum_v A_v-\sum_p {\tilde{B}_p}+\textrm{h.c.}
\end{equation}
Here, $A_v$ is simply the vertex constraint enforcing the string-net branching rules, which acts as the product of Pauli $Z$ operators on the four links around vertex $v$. $\tilde{B}_p$ is a plaquette term associated to the $xy$ oriented plaquette $p$, and has the form
\begin{equation}
    \tilde{B}_p=X_1X_2X_3X_4\phi_p(\{Z\})P_p
\end{equation}
where edges 1, 2, 3, and 4 bound plaquette $p$, $\phi_p(\{Z\})$ is a particular phase-valued function of the Pauli $Z$ variables in the vicinity of plaquette $p$, and $P_p$ is a projector onto the subspace satisfying the vertex constraints in the vicinity of $p$. The function $\phi_p$ takes the same form for all plaquettes in the same plane $z_0$, but may vary between planes if the model is not translation invariant in the $z$ direction. The particular form of $\phi_p$ depends on the element of the cohomology group $H^3[\mathbb{Z}_2^L,U(1)]$ represented by the model. It has been shown that representative models of all cohomology classes admit Hamiltonian descriptions of this form.~\cite{GeneralizedStringNet}

We now consider coupling together three mutually perpendicular 1-foliated $\mathbb{Z}_2^L$ gauge theories oriented along the $xy$, $yz$ and $zx$ planes, via a $ZZ$ coupling between the two qubits 
on each link of the cubic lattice. In the strong-coupling limit, the two qubit degrees of freedom merge into one effective qubit, and the following effective Hamiltonian emerges:
\begin{equation}
    H_\textrm{3-fol}=-\sum_v (A_v^{xy}+A_v^{yz}+A_v^{zx})-\sum_c \tilde{B}_c
\end{equation}
This Hamiltonian is identical to $H_{XC}$, except that the cube operator takes the form
\begin{equation}
    \tilde{B}_c=\left(\prod_{e\in c}X_e\right)\phi_c(\{Z\})P_c
\end{equation}
where $\phi_c(\{Z\})$ is a phase-valued function of the Pauli $Z$ variables in the vicinity of cube $c$, and $P_c$ is a projector onto the subspace satisfing all vertex contraints in the vicinity of $c$. The function $\phi_c(\{Z\})$ is defined in terms of the phase functions $\phi_p(\{Z\})$ of the 1-foliated gauge theories as follows:
\begin{equation}
    \phi_c(\{Z\})=\prod_{p\in c} \phi_p(\{Z\}),
    \label{eq:fcfp}
\end{equation}
where the $\phi_p(\{Z\})$ are now interpreted to act on the merged effective qubits.

If we ungauge $H_\textrm{3-fol}$ by regarding the fractonic excitations of the cube terms $\tilde{B}_c$ as gauge charge, the resulting model contains one spin per site $\ve{r}$ of the dual cubic lattice. The Hamiltonian takes the form
\begin{equation}
    H^\textrm{un}_\textrm{3-fol}=-\sum_\ve{r}X_\ve{r}\phi_\ve{r}(\{Z\})
\end{equation}
where $\phi_\ve{r}(\{Z\})$ is the ungauged version of $\phi_c(\{Z\})$ for site $\ve{r}$ dual to cube $c$, which is a function of plaquette variables $ZZZZ$ (tensor product over the four edges in a plaquette) hence manifestly symmetric.
Because $\phi_c(\{Z\})$ has the form \ref{eq:fcfp}, it follows that $\phi_\ve{r}(\{Z\})$ has the form
\begin{equation}
\begin{split}
    \phi_\ve{r}(\{Z\})&=\phi_{\ve{r}+\ve{x}/2}(\{d_{\ve{r}+\ve{x}/2}\})\phi_{\ve{r}-\ve{x}/2}(\{d_{\ve{r}-\ve{x}/2}\})\\
    &\times\phi_{\ve{r}+\ve{y}/2}(\{d_{\ve{r}+\ve{y}/2}\})\phi_{\ve{r}-\ve{y}/2}(\{d_{\ve{r}-\ve{y}/2}\})\\
    &\times\phi_{\ve{r}+\ve{z}/2}(\{d_{\ve{r}+\ve{z}/2}\})\phi_{\ve{r}-\ve{z}/2}(\{d_{\ve{r}-\ve{z}/2}\})
\end{split}
\end{equation}
where each function $\phi_{\ve{r}+\ve{\sigma}/2}(d_{\ve{r}+\ve{\sigma}/2})$ is dual to one of the plaquette phase functions $\phi_p(\{Z\})$, and the variables $d_{\ve{r}+\ve{\sigma}/2}$, similar to the variables $d_\ve{r}$ in the main text, are defined as $d_{\ve{r}+\ve{\sigma}/2}=Z_\ve{r}Z_{\ve{r}+\ve{\sigma}}$ for $\ve{\sigma}=\ve{x},\ve{y},\ve{z}$ the three unit vectors. Hence, it is straightforwardly verified that $H^\textrm{un}_\textrm{3-fol}$ is the result of stacking three 1-foliated SSPTs with respective Hamiltonians
\begin{equation}
\begin{split}
    H^{xy}_\textrm{1-fol}=-\sum_\ve{r}X_\ve{r}\phi_{\ve{r}+\ve{z}/2}(\{d_{\ve{r}+\ve{z}/2}\})\phi_{\ve{r}-\ve{z}/2}(\{d_{\ve{r}-\ve{z}/2}\})\\
    H^{yz}_\textrm{1-fol}=-\sum_\ve{r}X_\ve{r}\phi_{\ve{r}+\ve{x}/2}(\{d_{\ve{r}+\ve{x}/2}\})\phi_{\ve{r}-\ve{x}/2}(\{d_{\ve{r}-\ve{x}/2}\})\\
    H^{zx}_\textrm{1-fol}=-\sum_\ve{r}X_\ve{r}\phi_{\ve{r}+\ve{y}/2}(\{d_{\ve{r}+\ve{y}/2}\})\phi_{\ve{r}-\ve{y}/2}(\{d_{\ve{r}-\ve{y}/2}\})
\end{split}
\end{equation}
Therefore, we have demonstrated that the Hamiltonian $H_\textrm{3-fol}$, obtained by strongly coupling three mutually perpendicular 1-foliated gauge theories, is dual to a weak SSPT Hamiltonian $H^\textrm{un}_\textrm{3-fol}$. Moreover, any weak 3-foliated SSPT can be constructed in this way, since $H^\textrm{un}_\textrm{3-fol}$ describes a stacking of three 1-foliated SSPTs which can in principle by arbitrary. While our discussion has focused on the $G=\mathbb{Z}_2$ case for simplicity, it can be straightforwardly generalized to arbitrary abelian group $G$.


\subsection{Condensation transitions}

As alluded to in the main text, and discussed in Ref.~\onlinecite{foliated5}, the procedure of stacking a 2D SPT onto a 3-foliated 3D SSPT is dual to the procedure of adding a 2D twisted gauge theory to a 3D twisted X-cube model, and condensing composite planon excitations composed of fracton dipole and 2D gauge charge pairs. This planon condensation process has the effect of confining the lineon dipoles and 2D gauge fluxes that braid non-trivially with these fracton dipoles and 2D gauge charges respectively, leaving deconfined only the composites of lineon dipoles and 2D gauge fluxes, which become the lineon dipoles of the condensed phase. The result is that the statistics of these lineon dipoles are now modified by the addition of the 2D gauge flux statistics.

Let us consider a simple example. Consider stacking an $xy$ oriented 2D $\mathbb{Z}_2$ SPT, with $\matr{M}$ matrix a single entry-matrix equal to 2, between layers $z=0$ and $z=1$ of a trivial 3D 3-foliated $\mathbb{Z}_2$ planar SSPT, dual to a copy of the X-cube model. This procedure is dual to adding a 2D double semion layer, whose gauge flux has semionic exchange statistics, and condensing the planon composed of the 2D gauge charge plus the fracton dipole centered around $z=1/2$. This condensation has the effect of confining both the 2D gauge flux and the lineon dipoles centered around $z=0$ and $z=1$, and leaving deconfined the composite of the $z=0$ lineon dipole and the 2D gauge flux, and the composite of the $z=1$ lineon dipole and the 2D gauge flux. Both of these composites therefore obtain semionic exchange statistics. This procedure corresponds to the addition of a single self-loop to a 3-foliated SSPT in our graphical notation (see Fig. 1).

This dual picture interpretation of the stacking construction of weak 3-foliated SSPTs sheds light on the correspondence with $p$-string condensation. The key point is that the $p$-string condensation procedure resulting in untwisted and twisted X-cube models, and the planon condensation procedure outlined above, commute with one another because they both involve condensation of pure gauge charge.
Therefore, one can construct the dual phases of weak 3-foliated SSPTs by 1) starting with three decoupled stacks of 2D toric code layers, 2) adding 2D twisted gauge theory layers and identifying the added gauge symmetries with existing gauge symmetries by condensing pairs of gauge charges, and 3) driving a $p$-string condensation transition. Since all 1-foliated SSPTs are weak, and thus can be constructed by stacking 2D SPTs, step 2 allows for the creation of arbitrary 1-foliated gauge theories. Therefore, any fracton model which is obtained by performing $p$-string condensation on intersecting 1-foliated gauge theories, is dual to a weak 3-foliated SSPT.

\end{document}